\documentclass[a4paper,11pt]{article}
\pdfoutput=1 

\usepackage{jcappub} 

\usepackage[T1]{fontenc} 
\usepackage{wrapfig}

\usepackage{cleveref}
\usepackage{natbib}
\usepackage{amssymb}
\usepackage{amsmath}
\usepackage{siunitx}
\usepackage{physics}
\usepackage{bm}
\usepackage[normalem]{ulem}
\usepackage{graphicx,xcolor}
\usepackage{booktabs}
\usepackage{lscape} 

\usepackage{verbatim}

\title{\mbox{Searching for Exploding Black Holes}}

\author[a,b]{Xavier Boluna,}

\author[a,b]{Stefano Profumo,}

\author[b,c,d]{Juliette Bl\'e,}

\author[b,e,f]{and Dana Hennings}

\affiliation[a]{\ignorespaces
	Department of Physics, University of California, Santa Cruz
	Santa Cruz, CA 95064, USA
}
\affiliation[b]{\ignorespaces
	Santa Cruz Institute for Particle Physics,
	Santa Cruz, CA 95064, USA
}
\affiliation[c]{\ignorespaces
	Universit\'e Paul Sabatier,
	118 Route de Narbonne, F-31062 Toulouse, France
}
\affiliation[d]{\ignorespaces
	Universit\'e Grenoble Alpes,
	621 Avenue Centrale, F-38400 Saint Martin d'Hérès, France
}
\affiliation[e]{\ignorespaces
	The Peddie School,
	201 South Main Street
Hightstown, NJ 08520, USA
}
\affiliation[f]{\ignorespaces
	Wesleyan University,
	45 Wyllys Ave, 
Middletown, CT 06459, USA
}

\emailAdd{profumo@ucsc.edu}

\abstract{The observation of the final stages of the evaporation of a light black hole, which Hawking referred to as ``black hole explosion'', would offer critical insights on quantum gravity and high-energy physics phenomena. Here, we explore, review, and revisit the observational features and rates expected for nearby, light, evaporating black holes, and we assess and compare the expected sensitivity of a broad range of observatories. We then focus on the search for candidate black hole explosions in archival data from the Fermi Large Area Telescope and Gamma-ray Burst Monitor, and outline possible future observational campaigns.
}

\begin{document}
\maketitle
\flushbottom

\section{Introduction}
\label{sec:introduction}
While the identity of the cosmological dark matter (DM) is still unknown, its role in shaping the universe as we observe it both at early and late times has now come into sharp definition \cite{ParticleDataGroup:2022pth}. In parallel, the quest for a consistent quantum theory of gravity is far from over (see e.g. \cite{Addazi:2021xuf} and references therein). Notably, at the intersection of these two major, open problems potentially lie light black holes of non-stellar origin, or Primordial Black Holes (PBHs, see e.g. \cite{Auffinger:2022khh,Carr:2020gox,Bird:2022wvk} for recent reviews of formation mechanism, constraints, and search strategies). The observation of the final stages of black hole evaporation would provide clues on the number of dark sector degrees of freedom independent of them being secluded from the visible sector \cite{Ukwatta:2015iba, Baker:2021btk, Baker:2022rkn}, and would shed light on physics at energy scales close to the Planck scale (and whether the Planck scale itself corresponds to its standard value, unlike in several beyond-the-Standard-Model extensions), thus offering insights on quantum gravity (see e.g. \cite{Lehmann:2019zgt} and references therein).

As envisioned by Hawking almost a half century ago \cite{Hawking:1974rv,Hawking_Carr}, simple physics relating the temperature and mass of black holes determines the runaway process of black hole ``explosions'' (to use the same expression Hawking introduced in Ref.~\cite{Hawking:1974rv}) and sets the universal luminosity of such events. In fact, both the luminosity and the {\em light-curve} (the photon emission rate as a function of time) of black hole evaporation are entirely fixed by the mass spectrum, charge, and spin of the degrees of freedom the hole can evaporate in, and by the photon yield associated with those same degrees of freedom. In the Standard Model (SM), this luminosity and light-curve are known with great precision, allowing for a sharp answer to the question of whether black hole explosions are detectable with current technology \cite{Auffinger:2022khh}.


Broadly, a PBH explosion would resemble a short gamma-ray burst. However, key differences exist:
\begin{itemize}
\item[(i)] the luminosity of an exploding PBH increases following a completely {\em universal} light-curve, and {\em never decreases}, at least assuming the existence of no additional degrees of freedom to the SM;
\item[(ii)] the spectrum of the gamma-ray emission, under the same assumption, is also universal;
\item[(iii)] PBH explosions must be {\em local}, i.e. within, at most, a few parsecs of Earth, to be detectable by current and planned facilities \cite{Ukwatta:2010zn}; the evaporation process thus potentially (albeit not necessarily) exhibits {\em proper motion}, depending on the age, distance, and PBH direction of motion relative the the line of sight;
\item[(iv)] finally, the intrinsic luminosity of exploding PBHs is exactly the same for every event.
\end{itemize}

The study we present here is not the first attempt at searching for PBH explosions with gamma-ray observations. The rate of explosions, on different time-scales and energy, has been constrained by direct observations with a number of different telescopes: in approximate order of appearance, limits have been obtained by the H.E.S.S. array of Cherenkov telescopes \cite{Glicenstein:2013vha} (see also the recent update in Ref.~\cite{HESS:2021rto}), MILAGRO \cite{Abdo:2014apa}, the Fermi Large Area Telescope (LAT) \cite{Fermi-LAT:2018pfs}, the VERITAS Cherenkov telescope array \cite{Archambault:2017asc}, and the water Cherenkov detector HAWC \cite{HAWC:2019wla}. The latter, at present, produced the most stringent constraints on the local ``explosion rate'', at $\dot n<3400\ {\rm pc}^{-3}\ {\rm yr}^{-1}$ at 99\% C.L.. Cline and collaborators have scoured early GRB data sets for events that could be potentially associated with PBH explosions, often, however, in the context of Hagedorn-type models that are now in conflict with experiments \cite{Cline:1995sz, Cline:1995jf, Cline:1996zg, Cline:1998fx,Cline:2011ci}. Ukawatta et al \cite{Ukwatta:2015mfb}  considered the prospects for utilizing a {\em network} of widely separated non-imaging space-based detectors (``inter-planetary network'', IPN) that could determine if a high-energy photon burst is at cosmological distances or whether it is nearby; intriguingly, several putative events, detected by three or more IPN spacecrafts have been associated with possibly nearby events \cite{Ukwatta:2015mfb}.

The remainder of this paper is structured as follows: In the next section \ref{sec:photons} we review the photon spectrum and light-curve of exploding PBH (note that other channels of PBH evaporation can be in principle used for detection, including notably cosmic rays \cite{Carr:1976zz, Korwar:2023kpy}, neutrinos \cite{Halzen:1995hu}, and gravitational waves \cite{Ejlli:2019bqj, Ireland:2023avg}); the following sec.~\ref{sec:rates} discusses  theoretical expectations for the density and rate of PBH explosions compatible with all other observational constraints, and presents a compact formula for the rate of explosions per unit volume per unit time today $\dot n_{\rm PBH}$ for a given generic {\em initial} PBH mass function $\psi_i$, in the form $\dot n_{\rm PBH}=\rho_{\rm DM}\psi_i(M_U)/(3t_U)$, where $M_U\simeq 5.1\times 10^{14}$ grams is the mass corresponding to a PBH with a lifetime equal to the age of the universe today, $t_U$. Sec.~\ref{sec:sens} discusses the sensitivity estimate for a PBH at a given distance and time to complete evaporation for a number of observatories. Our results of searching for candidate PBH explosion in long-term transient sources and gamma-ray burst catalogues are presented in sec.~\ref{sec:grb}, and, finally, our discussion, outlook, and conclusions in sec.~\ref{sec:discussion}.

\section{Photon emission from exploding black holes}\label{sec:photons}

The instantaneous and integrated photon emission from black holes of a given mass, charge, and spin is by now well understood and readily available via publicly accessible numerical codes such as {\tt BlackHawk} \cite{BlackHawk1}, that we utilize herein. The production of photons stems both from direct evaporation (``primary'' photons) and from the indirect production of photons from radiation or decay of different evaporation products (``secondary'' photons). A particle species $i$ of mass $m$ spin $s$, angular momentum quantum numbers $(l,m)$, charge $q$ is emitted by a black hole (BH) (which we take as having a mass $M$, charge $Q$ and angular momentum $\Omega$) with energy $E$ at a rate\footnote{Note that the particle energy may be corrected by the effective chemical potentials due to spin or charge coupling \cite{Auffinger:2022khh}.} 
\begin{equation}\label{eq:production}
    \frac{{\rm d}^2N_i}{{\rm d}E\ {\rm d}t}(E)=\frac{1}{2\pi}\frac{\Gamma_{E,s,q,l,m}(M,Q,\Omega,...)}{e^{E/T_{\rm BH}}-(-1)^{2s}}\theta(E-m),
\end{equation}
where $T_{\rm BH}=\kappa/2\pi$ is the BH temperature ($\kappa$ is the surface gravity of the BH, and $T_{\rm BH}=1/(8\pi M_{\rm BH})$, in natural units, for uncharged, non-spinning Schwarzschild black holes), and $\Gamma$ are the so-called ``greybody factors'', that depend on both the particle and the BH quantum numbers, mass, and energy \cite{MacGibbon:1990zk}. The flux of secondary particles $j$ is computed from Eq.~(\ref{eq:production}) as
\begin{equation}
\frac{{\rm d}^2N_j}{{\rm d}E\ {\rm d}t}(E)=\int_0^\infty\sum_i {\rm BR}_{i\to j}(E,E^\prime)\frac{{\rm d}^2N_i}{{\rm d}E^\prime\ {\rm d}t}(E^\prime){\rm d}E^\prime,
\end{equation}
where ${\rm BR}_{i\to j}(E,E^\prime)$ is the particle physics branching ratio describing how a particle $j$ of energy $E$ is eventually produced from particle $i$ of energy $E^\prime$.

In the Standard Model, the approximate photon emission per unit time from a BH of temperature $T_{\rm BH}$, integrated over the energy range 0.3 GeV to 100 GeV, is, at late times, \cite{MacGibbon:1990zk}
\begin{equation}\label{eq:ngammadot}
    \dot N_\gamma\simeq 1.4\times 10^{29}\left(\frac{T_{\rm BH}}{\rm TeV}\right)^{1.6}{\rm sec}^{-1},
\end{equation}
which translates into a photon flux, at a distance $d$, of
\begin{equation}
    \phi_\gamma\simeq 1.2\times 10^{-9}\left(\frac{\rm pc}{d}\right)^2\left(\frac{T_{\rm BH}}{\rm TeV}\right)^{1.6}{\rm sec}^{-1}{\rm  cm}^{-2};
\end{equation}
As a function of the black hole's mass $M_{\rm BH}$, this, in turn, reads
\begin{equation}
    \phi_\gamma\simeq 1.1\times 10^{-9}\left(\frac{\rm pc}{d}\right)^2\left(\frac{10^{10}\ {\rm g}}{M_{\rm BH}}\right)^{1.6}{\rm sec}^{-1}{\rm  cm}^{-2}.
\end{equation}
For a non-spinning, uncharged ``Schwarzschild'' black hole (for which the formulae above apply), the mass evolution of the black hole with time can be inferred from the emission rates above and from the temperature-mass relation, and can be schematically cast in the form
\begin{equation}\label{eq:massev}
    \frac{dM_{\rm BH}}{dt}=-\frac{\alpha(M_{\rm BH})}{M_{\rm BH}^2,}
\end{equation}
with the ``Page coefficient'' $\alpha(M)$ dependent on the particle degrees of freedom that the hole can evaporate to; as a result, and assuming that $\alpha(M)=\alpha$ is constant at large-enough temperatures, the lifetime of a light PBH goes as $\tau_{\rm BH}\sim M_{\rm BH}^3/(3\alpha)$. In particular, $\tau_{\rm BH}(M_U)=t_U$ for $M_U\simeq 5.1\times 10^{14}$ g.

\section{Density and explosion rates for evaporating black holes}\label{sec:rates}

Constraints on the relative contribution of PBHs to the dark matter mass density are usually cast in terms of the maximal fraction $f_{\rm max}(m)$ of the cold dark matter density that can consist of PBH of a given, single mass $m$. Constraints on the abundance of PBHs of a given mass stem from a number of observations and considerations (for a comprehensive recent review, see \cite{Carr:2020gox}): the abundance of PBHs with masses such that they have evaporated in the very early universe are constrained by the abundance of light elements (which would be perturbed by the high-energy emission from PBH evaporation), specifically and most strongly between $10^9$ grams and $10^{13}$ grams, masses corresponding to lifetimes around the beginning and the end of Big Bang Nucleosynthesis \cite{Harada:2001kc}; by spectral distortions of the CMB, and by disruption of the CMB anisotropy power spectrum \cite{Acharya:2020jbv}, and, finally, by excessively contributing to the extra-galactic background light \cite{Harada:2001kc}. For masses above $10^{14}$ grams, black hole evaporation is ongoing, and constrained by direct observation of the products of evaporation (cosmic rays and gamma rays); up to $10^{17}$ grams, the strongest constraints are from evaporation to gamma-rays or to positrons (see e.g. the recent \cite{Korwar:2023kpy} that produced the strongest constraints from evaporation); in between $10^{17}$ and $10^{23}$, PBH can be (but cannot exceed) the entirety of the cosmological cold dark matter; in the large range between $10^{23}$ and $10^{35}$ grams, the strongest constraints arise from microlensing surveys from different telescopes and observational strategies \cite{Smyth:2019whb, DeRocco:2023gde}; finally, around stellar masses, strong constraints arise from gravitational wave production \cite{Vaskonen:2019jpv}, from the effects on the CMB of relativistic electrons produced by accretion on stellar-mass PBH \cite{Serpico:2020ehh}, and from the stability of stellar clusters such as the one at the center of the Eridanus II dwarf galaxy \cite{Brandt:2016aco}. For a summary of constraints, the Reader is directed to fig.~4 and fig.~10 of Ref.~\cite{Carr:2020gox}: In what follows, we utilize the constraints shown in those figures.

The (maximal possible) number density of monochromatic PBHs at a mass $m_{\rm PBH}$ can be simply cast, in terms of the local dark matter density $\rho_{\rm DM}(\vec r)$, as: 
\begin{equation}
    n_{\rm PBH}(\vec r)=f_{\rm max}(m_{\rm PBH})\frac{\rho_{\rm DM}(\vec r)}{m_{\rm PBH}}.
\end{equation} For non-monochromatic mass functions, say for a mass function $\psi(M)\propto M\frac{dn_{\rm PBH}}{dM}$, normalized as customary to $f_{\rm PBH}$, i.e.
\begin{equation}
    \int dM\psi(M)=f_{\rm PBH},
\end{equation}
the equivalent constraints can be cast as \cite{Carr:2017jsz}
\begin{equation}\label{eq:constr}
    \int dM\frac{\psi(M)}{f_{\rm max}(M)}<1.
\end{equation}
Usually, $\psi(M)=\psi_i(M)$ indicates the mass function at PBH creation. One then needs to time-evolve the mass function in the standard way \cite{Carr:2017jsz}, obtaining a final mass function $\psi_f$ (see also \cite{Halzen:1991uw})
\begin{equation}
   \frac{\psi_f(M)}{M}= \frac{dn}{dM}= \frac{dn}{dM_i} \frac{dM_i}{dM},
\end{equation}
where, from Eq.~(\ref{eq:massev}), for slowly-varying $\alpha(M)$,  
\begin{equation}
    M_i^{3}=M^3+3\alpha(M)t,\qquad \frac{dn}{dM_i} =\frac{\psi_i(M_i)}{M_i}.
\end{equation}
Assuming that the Page coefficient $\alpha(M)=\alpha$ does not vary appreciably in the mass range of interest (more precisely, that $|d\alpha(M)/dM|t\ll M^2$, which is abundantly true in the standard case), one finds
\begin{equation}
    \frac{dn}{dM}=\psi_i\left(\left(M^3+3\alpha t\right)^{1/3}\right)\frac{M^2}{M^3+3\alpha t}.
\end{equation}
Thus, the rate of explosions today, i.e. the rate at which the mass function crosses $M=0$ per unit volume per unit time corresponds to
\begin{equation}
    \dot n_{\rm PBH}=\lim_{t\to t_U}\lim_{M\to0}\frac{dn}{dM}\left(-\frac{dM}{dt}\right)=\lim_{t\to t_U}\lim_{M\to0}\rho_{\rm DM}\psi_i\left(\left(M^3+3\alpha t\right)^{1/3}\right)\frac{M^2}{M^3+3\alpha t}\frac{\alpha}{M^2},
\end{equation}
which gives our final result
\begin{equation}\label{eq:rateeq}
 \dot n_{\rm PBH}=\rho_{\rm DM}\frac{\psi_i\left(M_U \right)}{3t_U},
\end{equation}
with $t_U$ the age of the universe, and $M_U=3\alpha(0)t_U\simeq 5.1\times 10^{14}$ g. 

\begin{figure}[!t]
    \centering
   \mbox{ \includegraphics[width=0.45\textwidth]{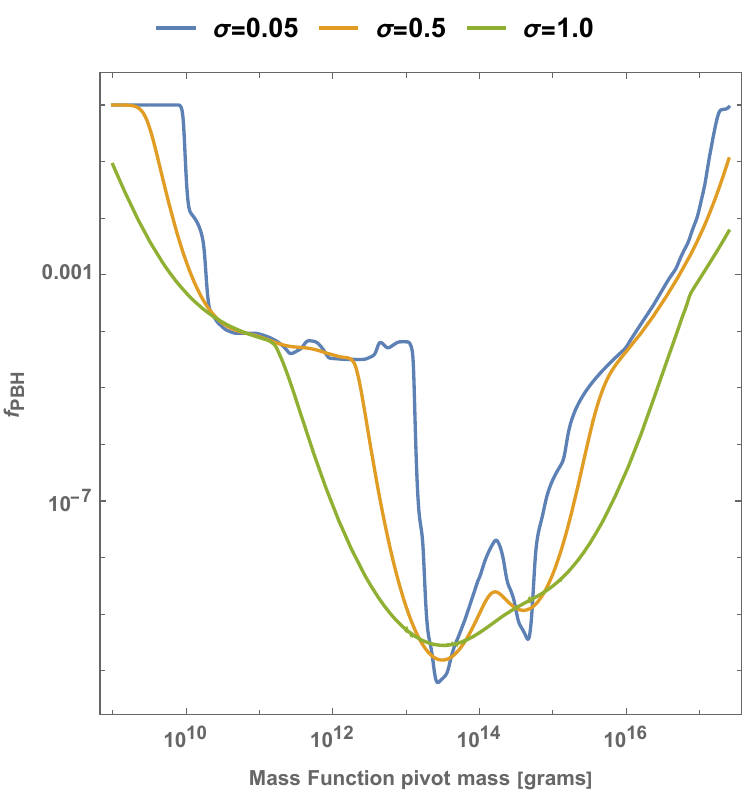}\qquad\ \includegraphics[width=0.46\textwidth]{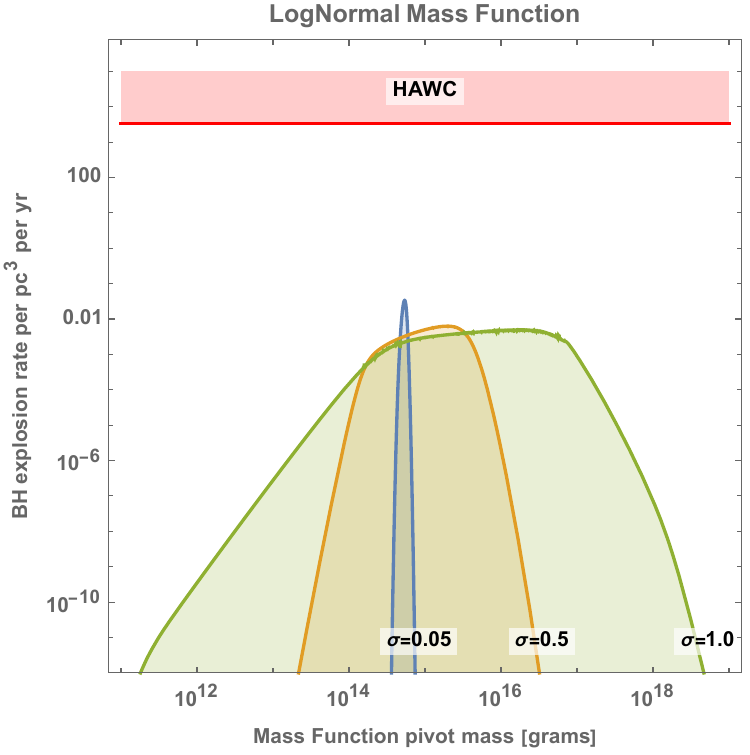}}
    \caption{Left: The constraints on $f_{\rm PBH}$ as a function of the pivot mass $M_*$ for the log-normal mass function of Eq.~(\ref{eq:lognormal}), for $\sigma=0.05,\ 0.5$ and 1.0; Right: the maximal PBH explosion rate per unit volume, as a unction of the pivot mass $M_*$, for the same choices of $\sigma$ as in the left panel.}
    \label{fig:lognormal}
\end{figure}

Let us now derive an upper limit to the rate in Eq.~(\ref{eq:rateeq}) for a few representative mass functions, with a normalization compatible with the constraints outlined and specified at the beginning of this section. Consider the following three instances of well-motivated mass functions:
\begin{enumerate}
\item {\bf Log-normal} mass function:
\begin{equation}\label{eq:lognormal}
    \psi(M,M_*,\sigma)=\frac{\exp\left(-\frac{\log\left(M/M_*\right)^2}{2\sigma^2}\right)}{\sqrt{2\pi}\sigma M}.
\end{equation}
Such mass function is the general expectation for the mass distribution of PBHs resulting from a smooth and symmetric peak in the inflationary power spectrum, as shown numerically in Ref.~\cite{Green:2016xgy}, and analytically in \cite{Kannike:2017bxn}, in the slow-roll approximation (see Ref.~\cite{Carr:2017jsz} for further details). 
Fig.~\ref{fig:lognormal}, left, shows the constraints, on the plane defined by the pivot mass $M_*$ and $f_{\rm PBH}$, resulting from three different choices of $\sigma=0.05,\ 0.5$ and 1.0, from Eq.~(\ref{eq:constr}): the broader the distribution, i.e. the larger $\sigma$ the stronger the limits from the larger ``tails'' leaking into regions of tighter constraints; for small $\sigma$, instead, the limits closely resemble the monochromatic limits on $f_{\rm PBH}$. The right panel, again as a function of the pivot mass $M_*$, shows the explosion rate today, Eq.~(\ref{eq:rateeq}), for the maximal possible $f_{\rm PBH}$ at that pivot mass, again for  $\sigma=0.05,\ 0.5$ and 1.0. The shaded regions are allowed by constraints on the PBH abundance for this particular mass function.

\begin{figure}[!t]
    \centering
   \mbox{ \includegraphics[width=0.45\textwidth]{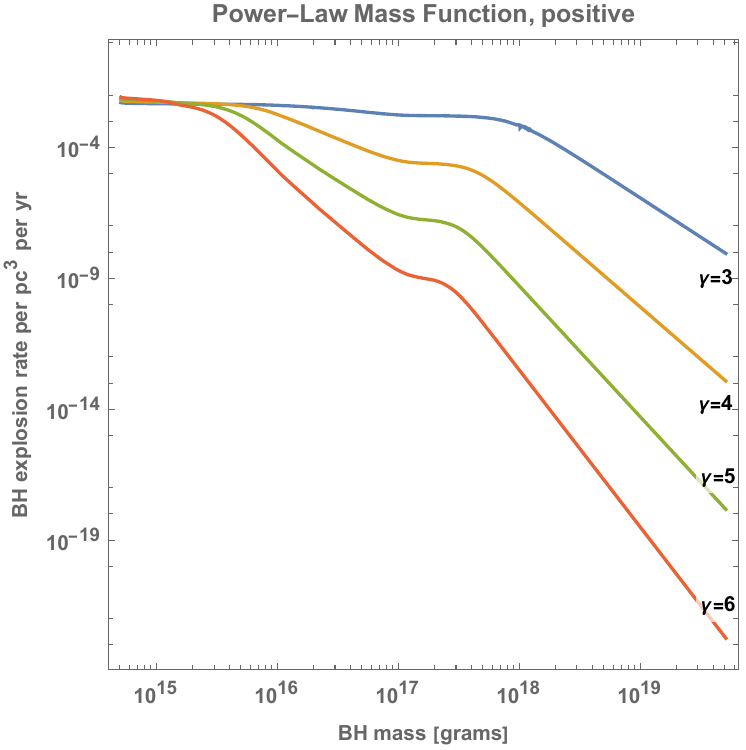}\qquad\ \includegraphics[width=0.45\textwidth]{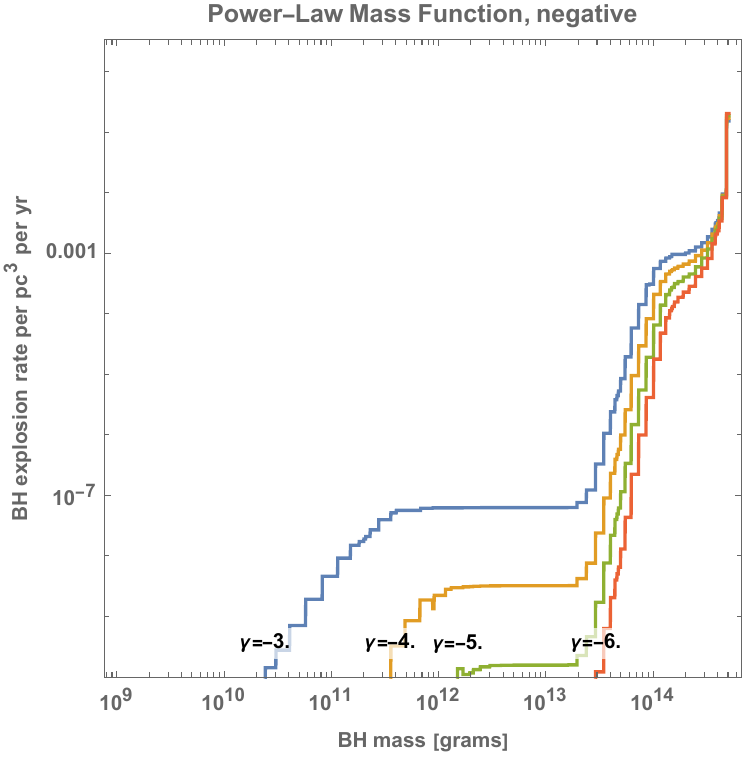}}
   \mbox{ \includegraphics[width=0.45\textwidth]{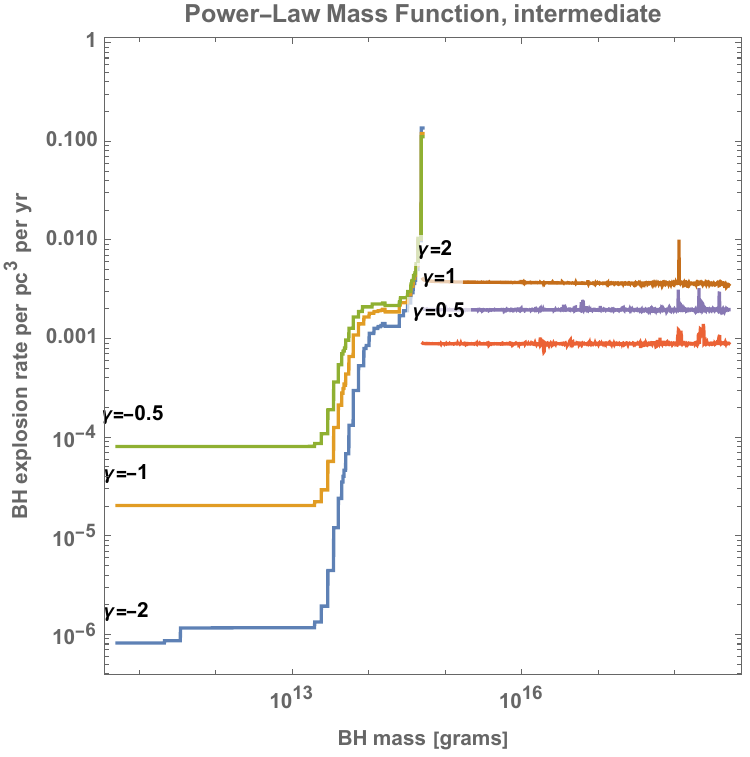}\qquad\ \includegraphics[width=0.45\textwidth]{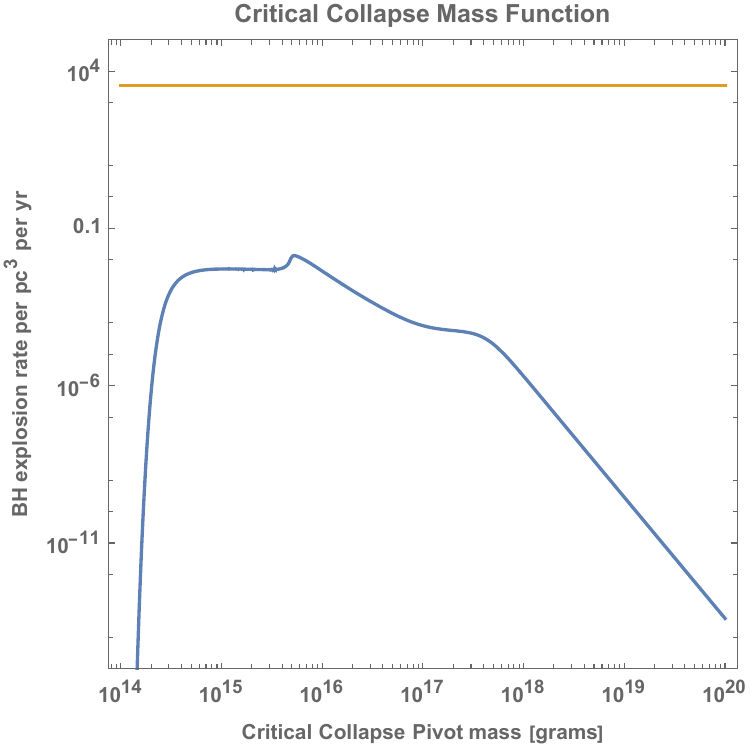}}
    \caption{The PBH evaporation rate per unit volume, $\dot n_{\rm PBH}$, for the power-law mass function of Eq.~(\ref{eq:powerlawp}-\ref{eq:powerlawp}), for large $|\gamma|\ge 3$ (top panels) and for intermediate $|\gamma|\le 2$ (bottom, left); bottom, right: same but for the critical collapse mass function of Eq.~(\ref{eq:cc}).}
    \label{fig:plcc}
\end{figure}

The peak appearing in fig.~\ref{fig:lognormal}, right, for small enough values of $\sigma$, is centered around $M_*=M_U$, so that the positive-definite argument of the exponential is 0; we find that 
\begin{equation}
\dot n_{\rm PBH}\simeq \frac{1.2\times 10^{-3}\ {\rm pc}^{-3}{\rm yr}^{-1}}{\sigma},
\end{equation}
implying that the current constraints from HAWC, at the level of 3400 ${\rm pc}^{-3}\ {\rm yr}^{-1}$ \cite{HAWC:2019wla},  for this particular mass function  constrain $\sigma\gtrsim 3.5\times 10^{-7}$. For smaller values of $\sigma$ the HAWC constraints {\em can} be violated while not violating the constraints on $f_{\rm PBH}$ at that given $M_*$ (in other words, for sufficiently small $\sigma$ the HAWC constraints are {\em the strongest} constraints on $f_{\rm PBH}$).

\item {\bf Power-law}: here, the mass function is defined as
\begin{equation}\label{eq:powerlawp}
    \psi(M,M_*,\gamma)=\frac{\gamma}{M}\left(\frac{M}{M_*}\right)^\gamma\ {\rm for}\ M<M_*,\ 0\ {\rm otherwise}\quad (\gamma>0)
\end{equation}
\begin{equation}\label{eq:powerlawn}
    \psi(M,M_*,\gamma)=-\frac{\gamma}{M}\left(\frac{M}{M_*}\right)^\gamma\ {\rm for}\ M>M_*,\ 0\ {\rm otherwise}\quad (\gamma<0).
\end{equation}
Power-law mass functions appear both if PBHs form from scale-invariant density fluctuations, or from the collapse of cosmic strings \cite{Carr:2017jsz}; remarkably, in both cases the power-law index $\gamma$ is uniquely related to the equation of state parameter $w$, where $p=w\rho$, when the PBHs form \cite{Carr:1975qj} via the relation 
\begin{equation}
    \gamma=-\frac{2w}{1+w}.
\end{equation}
Note that for a non-inflationary cosmology where $-1/3<w<1$, the natural range for $\gamma$ is $-1<\gamma<1$. Values outside this range are, however, generally possible in non-standard formation scenarios.

For a power-law mass function, evidently, the maximal explosion rate is achieved at $M_*=M_U$, and it corresponds to $\dot n_{\rm PBH}\sim 0.1\ {\rm pc}^{-3}{\rm yr}^{-1}$ for $\gamma<0$ and to $\dot n_{\rm PBH}\sim 0.01\ {\rm pc}^{-3}{\rm yr}^{-1}$ for $\gamma>0$. 
We show the maximal PBH explosion rate $\dot n_{\rm PBH}$ as a function of $M_*$ for positive (left) and negative (right) $|\gamma|\ge 3$ in the top panel of Fig.~\ref{fig:plcc}, for different values of $\gamma$. The bottom, left panel of the same figure shows again the explosion rate, for values of gamma in the intermediate $|\gamma|<2$ range. In no case do we find a rate $\dot n_{\rm PBH}$ anywhere close to the current HAWC constraints.

\item {\bf Critical Collapse}: 
\begin{equation}\label{eq:cc}
        \psi(M,M_c)\propto M^{2.85} \exp{-(M/M_c)^{2.85}}.
\end{equation}
This mass function is expected to occur when PBHs for from a sharply peaked feature in the density fluctuation power spectrum \cite{Yokoyama:1998xd,Niemeyer:1999ak,Musco:2012au,Carr:2016hva}. The resulting spectrum extends to arbitrarily low masses, with a cutoff corresponding to the mass $M_c$ associated with the horizon mass at the epoch of collapse \cite{Yokoyama:1998xd}. Note that
In the case of the single-parameter critical collapse mass function, we find that, after taking into account the constraints on $f_{\rm PBH}$, the absolute maximum rate corresponds to a pivot mass $M_c\simeq 5.7\times 10^{15}$ grams, and is approximately $\dot n_{\rm PBH}\sim 1.3\times 10^{-2}\ {\rm pc}^{-3}{\rm yr}^{-1}$ (see Fig.~\ref{fig:plcc}, bottom right) hence, again, far from the current HAWC constraints.

\end{enumerate}

In summary, for the three physically-motivated classes of mass functions we examined, the expected rate of PBH explosions is generically well below current observational limits, except for a log-normal mass function with an extremely narrow width and a pivot mass very close to the mass corresponding to a lifetime equal to the present age of the universe.

We note, however, and on a more positive note, that there exist scenarios where PBH can be injected at very late times, where $\dot n_{\rm PBH}$ then depends on the injection rate and mass function at injection, but rather on late-times physics. This class of scenarios were recently reviewed and investigated in detail in Ref.~\cite{Picker:2023lup, Picker:2023ybp} to where we refer the interested Reader.



\section{Sensitivity estimates and search strategies}\label{sec:sens}
\begin{wrapfigure}{r}{0.4\textwidth}
  \begin{center}
    \includegraphics[width=0.37\textwidth]{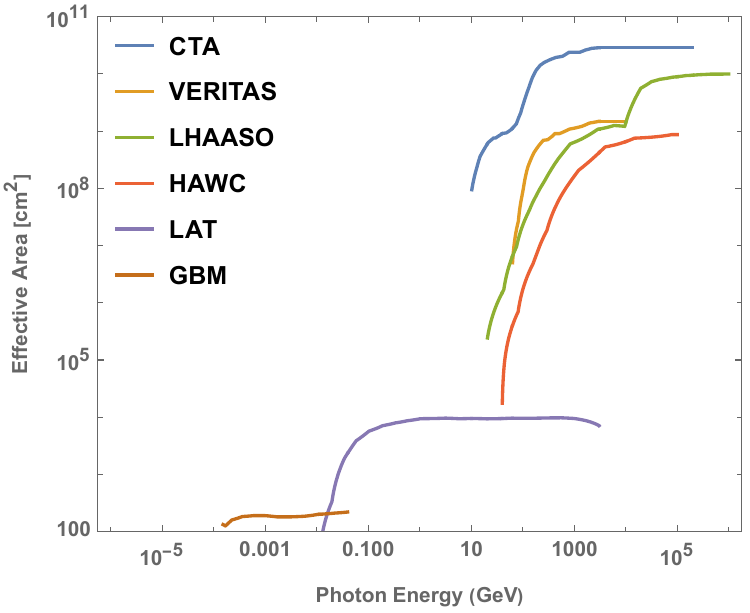}
  \end{center}
  \caption{Effective area of the observatories under investigation; outside the ranges where data were available, we conservatively set the effective area to zero.}\label{fig:effarea}
\end{wrapfigure}

We now proceed to compute updated estimates of the sensitivity of current and future gamma-ray detectors to exploding, light PBHs. To this end, we collected and digitized effective areas and angular resolution, as a function of energy, for the following detectors: GBM \cite{Meegan:2009qu}, LAT \cite{Fermi-LAT:2009ihh}, Veritas \cite{Weekes:2001pd}, HAWC \cite{HAWC:2011gts}, LHAASO \cite{LHAASO:2019qtb} and CTA \cite{CTAConsortium:2013ofs}; we show the effective areas we employ, as a function of energy, in fig.~\ref{fig:effarea} (note that at very high energy, where data are not available or the effective areas are not specified by the experimental collaborations, we conservatively set the effective areas to zero); for simplicity, we employ an energy-independent angular resolution of $\Delta\Omega=10^{-3}$ sr for the GBM and the LAT, of $\Delta\Omega=10^{-4}$ sr for LHAASO, and of $\Delta\Omega=10^{-5}$ sr for HAWC, VERITAS, and CTA.

As for the gamma-ray background, we digitized the isotropic extragalactic background as measured by the LAT \cite{Fermi-LAT:2010pat}, with an extrapolation at very high energy of 
\begin{equation}
    \phi_B(E)=1.4\times 10^{-6}(E/{\rm GeV})^{-2.1}\ {\rm cm}^{-2}{\rm GeV}^{-1}{\rm sec}^{-1}{\rm sr}^{-1}.
\end{equation}
We assume an optimal observing circumstance where the Galactic diffuse background is negligible compared to the extragalactic background (equivalently, we assume that the evaporating PBH is at high-enough Galactic latitude so that the Galactic diffuse emission is relatively negligible); for very high-energy gamma-ray telescopes we also add a background due to mis-identified cosmic-ray events, with a spectrum $d\phi_{\rm CR}/dE=0.95 (E/{\rm GeV})^-2.65\ {\rm cm}^{-2}\ {\rm sec}^{-1}\ {\rm sr}^{-1}$ \cite{Lipari:2019jmk}, with an instrumental rejection rate that we assume to be at the level of a contamination of 1 in 10$^5$ for all telescopes, independent of energy \cite{Weekes:2001pd, HAWC:2011gts, LHAASO:2019qtb, CTAConsortium:2013ofs}. For VERITAS and CTA we also included the cosmic-ray electron background, assumed to have a spectrum $d\phi_{\rm e}/dE=0.03 (E/{\rm GeV})^-3.2\ {\rm cm}^{-2}\ {\rm sec}^{-1}\ {\rm sr}^{-1}$, and with a rejection efficiency of 1 in 10$^2$ \cite{Staszak:2015kza} (this background is subdominant to the other two we consider).

Given a PBH evaporating at a distance $d$, we adopt the following procedure to define a {\em detection}: we ask that a given telescope of effective area $A_{\rm eff}(E)$ and angular resolution corresponding to a solid angle $\Delta\Omega(E)$ can detect\footnote{The signal to noise is always maximized for the smallest possible angular region, which we assume to correspond to the instrumental angular resolution at a given energy, as specified above.}, in a statistically significant manner, a signal over background. As described in sec.~\ref{sec:photons} above, PBHs emit both primary and secondary photons (the latter from the decay into photons of other species produced upon evaporation) at a differential rate that depends on the black hole mass, and thus on time, $d^2N_\gamma(M_{\rm BH}(t))/(dt\ dE)\ {\rm sec}^{-1}\ {\rm GeV}^{-1}$; assuming, as we do here, that the hole is spinless, the emission is isotropic, and the flux of photons at the detector is 
\begin{equation}
    \phi_\gamma(E)=\frac{d^2N}{dt\ dE}(M_{\rm BH}(t))\frac{1}{4\pi d^2}.
\end{equation}
The number of {\em signal} photons detected over an observation time $t_f-t_i$, of duration at most $T_{\rm obs}=1$ year, is
\begin{equation}\label{eq:NS}
N_S=\int dE\int_{t_i}^{t_f}dt\frac{d^2N}{dt\ dE}(M_{\rm BH}(t))\frac{A_{\rm eff}(E)}{4\pi d^2}.
\end{equation}
Note that the relation between the time-to-expiration $\tau$ (i.e. time to complete evaporation {\em at the beginning of the observation}), i.e. time till $T_{\rm  BH}\to\infty$) and $M_{\rm BH}$ is, at low-enough BH masses, \cite{Carr:2020gox}:
\begin{equation}\label{eq:timetoevap}
    \frac{\tau}{407\ {\rm sec}}\simeq\left(\frac{M_{\rm BH}}{10^{10}\ {\rm gm}}\right)^{3}.
\end{equation}
We need to specify the limits of integration in Eq.~(\ref{eq:NS}): while for the energy integral those are automatically taken care of by the $A_{\rm eff}(E)$, which goes to zero at very large and very small energies, for the time integral (where time refers to the black hole time-to-expiration) we have, for a maximal observation duration of time $T_{\rm obs}$ and for a black hole with a leftover lifetime-to-expiration $\tau$ at the beginning of the observation  and of {\em total} lifetime $\tau_{\rm BH}$:
\begin{equation}
t_i=\tau_{\rm BH}-\tau;\qquad t_f={\rm Min}\left(\tau_{\rm BH}-\tau+T_{\rm obs},\tau_{\rm BH}\right),
\end{equation}
meaning that the observation will last for an amount of time $\tau$ if $\tau<T_{\rm obs}$, i.e. if the observation time is longer than the remaining black hole lifetime, and it will last $T_{\rm obs}$, thus ending when the black hole's time to expiration is $\tau-T_{\rm obs}$, if $\tau>T_{\rm obs}$. The number of {\em background} photons corresponding to the observation, meanwhile, is
\begin{equation}
    N_B={\rm Min}\left(T_{\rm obs},\tau\right)\times \int dE\  \phi_B(E) A_{\rm eff}(E)\Delta\Omega(E).
\end{equation}
Note that we neglect (i) the dead-time fraction of detectors, and (ii) the dependence of the effective area and angular resolution on the angular direction  where the event occurs.

\begin{figure}[!t]
    \centering
    \includegraphics[width=\linewidth]{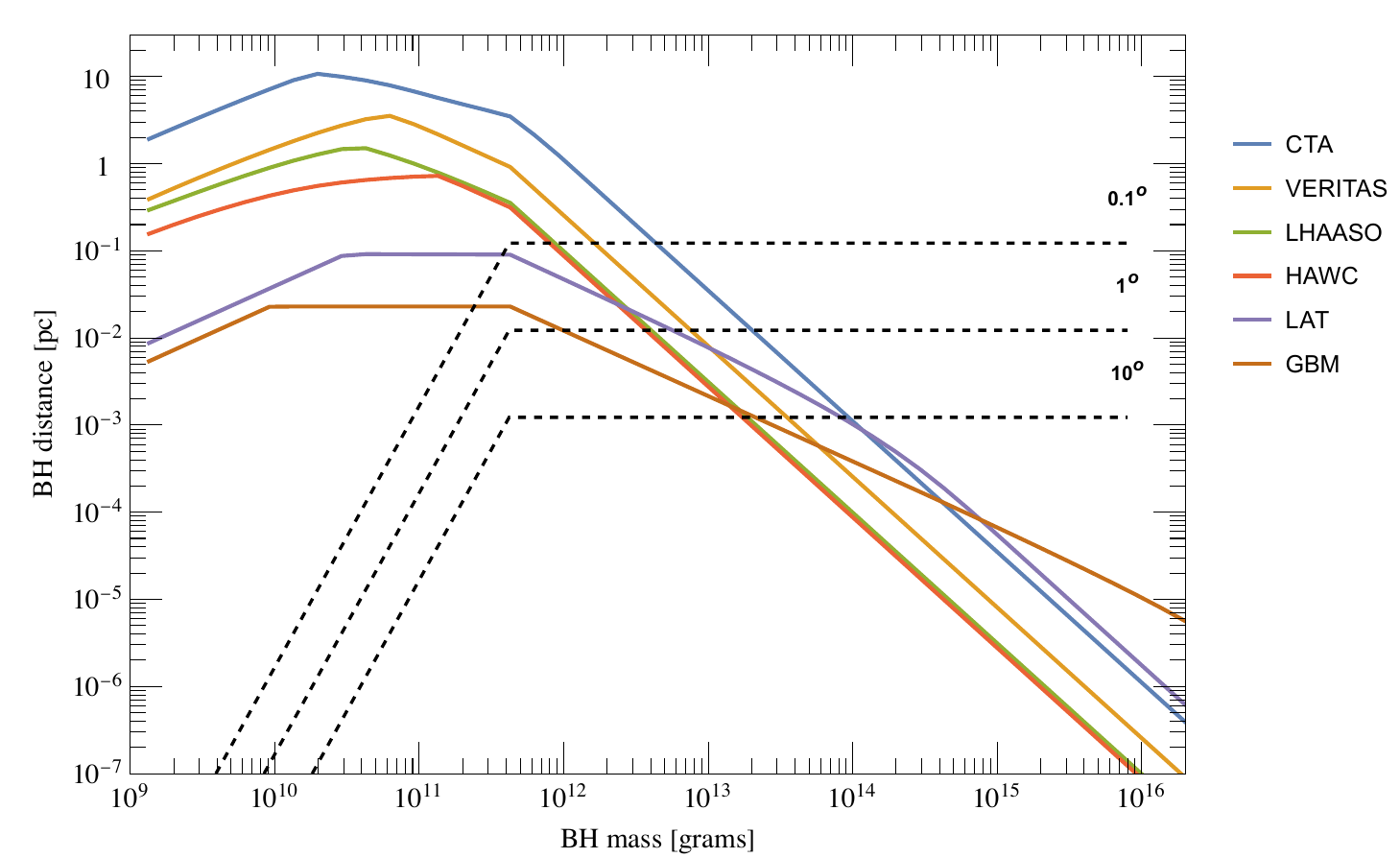}
    \caption{Maximum visible distance curves with respect to black hole mass for a number of modern gamma-ray detectors. The black dashed lines indicate the distance, at a given mass, at which the  proper motion during an observation up to one year long may exceed 0.1, 1 and 10 degrees.}
    \label{fig:detsens}
\end{figure}

We ask for two conditions for a detection of a signal over background:
\begin{enumerate}
\item $N_S>10$, i.e. to detect at least 10 signal photons;
\item $N_S/\sqrt{N_B}>5$, i.e. that the approximate significance of the signal to background be larger than 5 sigma.
\end{enumerate}
Adopting the formalism outlined above, we obtain the detector sensitivities shown in Fig.~\ref{fig:detsens}, in the plane defined by the black hole initial mass (or time to evaporation, related to the former by Eq. (\ref{eq:timetoevap})) versus distance. Anything {\em below} the curves shown should be detectable by the corresponding facility. 
We also show, with dashed lines, the contours corresponding to the maximal proper motion in the sky during the relevant observation time, ${\rm Min}\left(T_{\rm obs},\tau\right)$, assuming $T_{\rm obs}=1$ yr. Above the dashed lines, proper motion is {\em below} the angular size shown on the plot. Notice that the angular proper motion depends on the angle $\psi$ between the direction of motion of the PBH and the line of sigh direction; what we show corresponds to $\psi=90^\circ$, but for a generic $\psi$, the resulting angular proper motion is suppressed by a factor $\sin\psi$.

Our findings illustrate that at present the best detector to discover evaporating black holes with a time-to-expiration $\tau$ shorter than a few years (equivalently, masses below, roughly $10^{12}$ grams) are Cherenkov telescope arrays; rather surprisingly, we find that the VERITAS array outperforms both HAWC and the future LHAASO telescopes, as a result of the effective area of VERITAS exceeding that of the other detectors in the critical range between a few 100 GeV and 10 TeV; for more massive and longer-lived PBHs, with masses between, approximately, $10^{13}$ and $10^{15}$ grams, the best currently-operating telescope is the LAT; for even more massive PBHs, the best telescopes are the smaller but lower-energy detectors GBM and BATSE (not shown, but similar to the GBM). The future CTA facility will outperform current observatories up to very large masses/low temperatures. 

The facilities hierarchy we find depends primarily on a combination of the effective area of the telescopes, their angular resolution and background rejection capability, and on their energy range: the lighter the PBH, the higher the temperature etc, thus our results are not entirely unexpected. The specific shape of the sensitivity curves is also not unexpected, and it reflects, at short lifetimes/low masses (left), the time needed to accumulate a significant number of photons, while on the right there is mix of declining effective areas at lower energies, and a larger accumulated number of background photons. The most promising evaporating PBH for detection are in the few 10$^{10}$ grams range (thus, objects that evaporate on time scales of the order of a few minutes), and located, at most, at a distance of a few parsecs, or $\sim10^{19}$ m, or $\sim10^8$ AU.

Note that for masses above approximately $M_{\rm 1yr}\simeq 4\times 10^{11}$ grams, for which the remaining black hole lifetime is $\tau(M\gtrsim M_{\rm 1yr})>1$ year, the observation time $T_{\rm obs}$ is less than the remaining black hole lifetime, thus the {\em explosion} is not observed. For $M>M_{\rm 1yr}$ much less high-energy radiation is detected, limiting rapidly the performance of very high energy telescopes, while boosting the performance of lower energy telescopes (LAT, GBM). At large enough masses, the temperature of the black hole is low enough that so little high-energy radiation is emitted that the best detector, for such slowly evaporating holes, is the GBM (the turnover happens at masses $M\sim 10^{15}$ grams, corresponding to temperatures of around 10 MeV).

\begin{figure}[!t]
    \centering
    \includegraphics[width=0.65\linewidth]{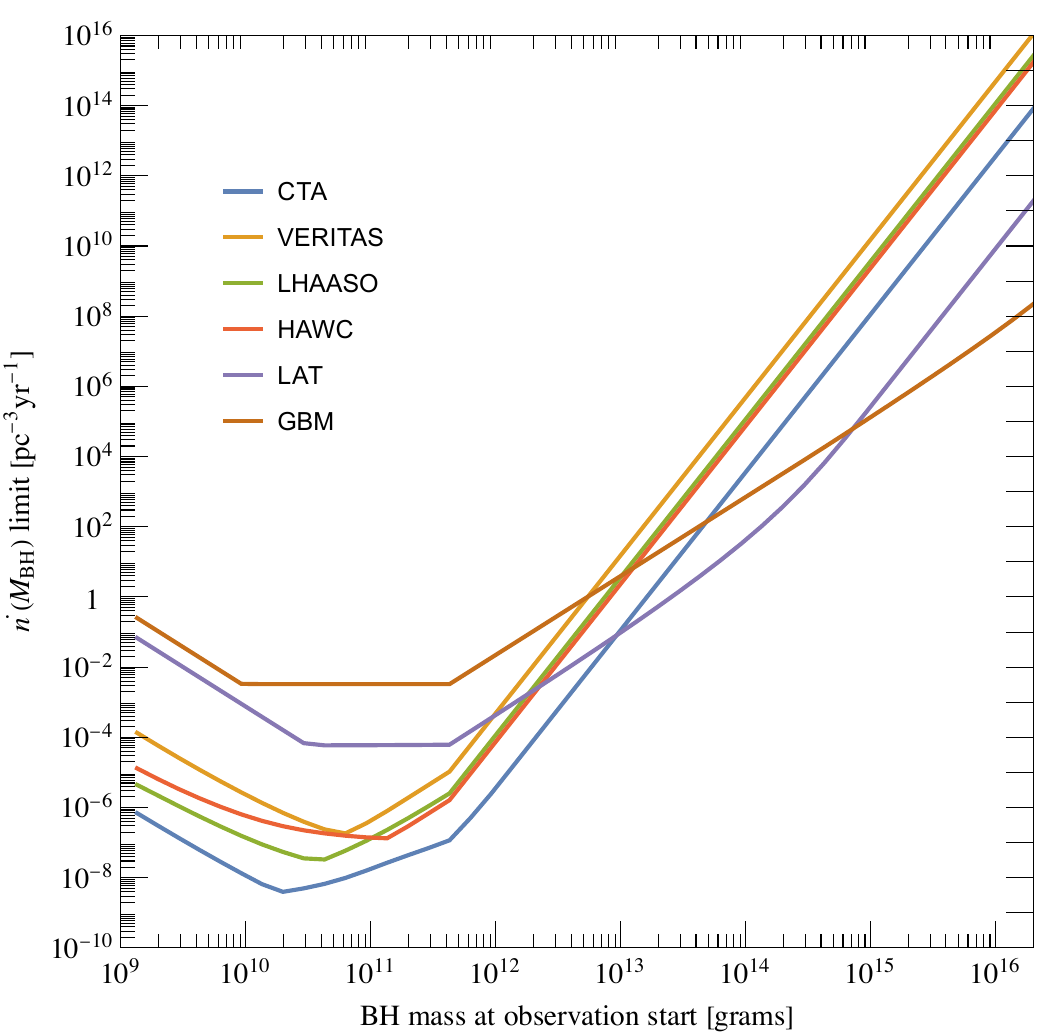}
    \caption{Projected constraints on the explosion rate $\dot n (M)$ as a function of the  BH mass at observation start.}
    \label{fig:rateconstraints}
\end{figure}
Given an observation of duration $T_{\rm obs}$ with a telescope sensitive to events out to a distance $d(M)$ as computed and shown in fig.~\ref{fig:detsens}, and of field of view $\Omega_{\rm fov}$, the resulting constraints on the PBH evaporation rate $\dot n (M)$ as a function of the initial BH mass are given by:
\begin{equation}
    \dot n(M)=\frac{3}{T_{\rm obs}\cdot d^3(M)\cdot \Omega_{\rm fov}}.
\end{equation}
Crudely approximating the field of view of the observatories under consideration as energy independent, and as 15\% of the sky for HAWC, 4.5 deg for CTA, 3.5 deg for VERITAS, 30 deg for LHAASO, 15\% of the sky for the LAT, and 2.4 sr for the GBM, we find the evaporation rate constraints shown in fig.~\ref{fig:rateconstraints}; we note that our findings, for $M\sim M_U$ are consistent with the limits from the HAWC collaboration \cite{HAWC:2019wla}. As expected, for short-duration events the observatories with the highest-energy sensitivity and the largest fields of view outperform other observatories; the LAT is optimal for masses in excess of 10$^{13}$ grams and up to 10$^{15}$ grams, when low-energy telescopes such as the GBM, also endowed with a large field of view, are the most constraining.

\subsection{Spectral energy index}

\begin{figure}[!t]
    \centering
    \begin{minipage}{\linewidth}
            \centering
            \includegraphics[width=.75\linewidth]{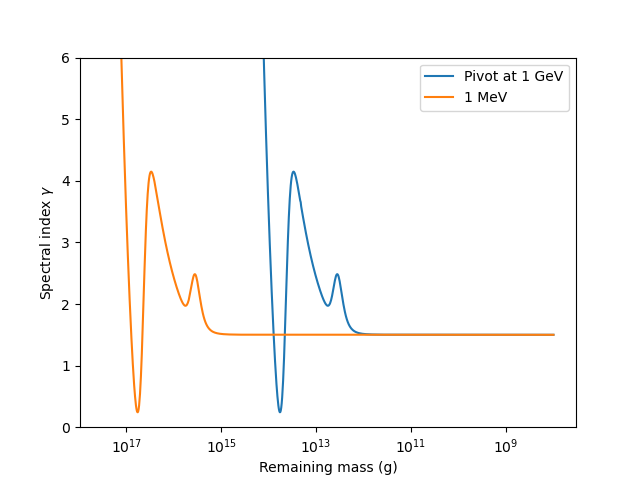}
            \caption{Spectral power-law index $\gamma$ at 1 GeV and 1 MeV as a function of the evaporating BH mass.}
            \label{fig:gammasp}
    \end{minipage}
\end{figure}
In addition to a universal light-curve, barring non-standard additional degrees of freedom, PBHs also feature a fully calculable, and universal, spectral evolution.  
%
In order to produce a quantity that is directly comparable to data, we compute, with the {\tt BlackHawk} code, the slope $\gamma$, at a fixed ``pivot energy'' $E_0$, of the {\em total differential photon} spectrum, assumed to have a power-law behavior $\phi(E)=dN_\gamma/dE\sim E^{-\gamma}$; that is, we calculate the quantity 
\begin{equation}\label{eq:spectralindex}
    \gamma(E_0)=-\lim_{E\to E_0}\frac{E}{\phi(E)}\frac{d\phi(E)}{dE}
\end{equation}
corresponding to an evaporating black hole of a given mass. 
We show in Fig.~\ref{fig:gammasp} the values of $\gamma$ when the pivot energy is fixed at $E_0=1$ GeV (blue line) and at $E_0=1$ MeV (orange line), as a function of the black hole mass. Notice that at sufficiently large mass, $\gamma$ grows very quickly due to the exponential decay in the emission spectrum at the pivot energy (this obviously happens at larger masses for lower pivot energies). Importantly, the asymptotic spectral index, at sufficiently small PBH mass, i.e. close to the endpoint of evaporation, is universal and independent of the pivot energy (i.e. independent of which energy one computes the slope of the spectrum at) at $\gamma(M_{\rm PBH}\to 0)\to 1.5$, which will inform the observational campaign presented below.


\subsection{The effect of proper motion}
Proper motion describes changes in the angular position of objects in the sky as they are observed from Earth, and is an important observational quantity for long-term, relatively dim, nearby sources. Many observatories in the gamma-ray domain lack capability to associate long-term sources over wide angular variations: For example, the \textit{Fermi}-LAT detection algorithm for transient sources considers sources with an error ellipse no greater than 15 degrees over the course of a decade of information (though observation windows for individual sources vary) \cite{LATTransients}.

We assume a PBH's average motion to be similar to the average Galactic dispersion $v \sim 220$ km s$^{-1}$ \cite{galacticDynamics}; the relevant time-scales correspond, as above, to Min$(\tau, T_{\rm{obs}})$. Here, $\tau$ is the remaining PBH lifetime and $T_{\rm{obs}}$ is the maximum telescope observation window (e.g. 10 years, with the \textit{Fermi}-LAT transient catalog). Then, for a PBH distance $d$, traveling at an angle $\psi$ with the line of sight,
\begin{equation}
    \theta = \frac{180^{\circ}}{\pi}v\times\frac{{\rm Min}(\tau, T_{\rm{obs}})}{d}\times \sin\psi
\end{equation}
At 0.01 parsecs, and for $\psi=90^\circ$, this corresponds to, roughly, 1 degree per year. This jumps beyond 10 degrees per year at 0.001 parsecs. Association of gamma-ray sources over large periods of time is generally subject to choice of algorithm 
and variance in the gamma-ray background \cite{Kunzweiler_2022}. We consider the conservative bound which excludes any sources with proper motion exceeding order $\geq$10 degrees per year, that would be hard to conclusively associate to the same source. We also verified that the dimming/brightening effect for $\psi\sim 0,\ 180^\circ$ is consistently negligible.

\section{Searches for black hole explosions with Fermi LAT and GBM data}\label{sec:grb}
The projections derived above paint a relatively consistent picture for the possible domains in which a PBH could be detected by modern gamma-ray detectors. In confronting publicly available data sets, we consider two general scenarios: the detection of the very last, runaway phase of the BH ``explosion'', and the steady evaporation phase that precedes it. While the former is generally associated with a phenomenology similar (albeit different in significant ways, as mentioned in the Introduction) to that of a gamma-ray burst (GRB), the latter would manifest itself as a transient source with {\em increasing} luminosity (barring the effect of proper motion, that could dim the object as it moves away from the observer which, as mentioned above, is however negligible for the distances under consideration), and with a universal light-curve and spectrum (again, barring the existence of significant dark-sector degrees of freedom).

\subsection{Long-term gamma-ray sources}\label{sec:longsrc}
In pursuit of long-term sources possibly associated with evaporating PBHs, we employ the LAT Transient Catalog \cite{LATTransients}. This source catalog uses data collected from August 4th, 2008 to August 15, 2018 -- over a decade of data. We focus our analysis on ``unnassociated'' sources, labelled as such for weak correlation with any other known sources at other wavelengths \cite{Fermi-LAT:2011sla}. At the time of retrieval from the HEASARC catalog\footnote{{\tt https://heasarc.gsfc.nasa.gov/W3Browse/fermi/fermiltrns.html}}, our query (based exclusively on the condition of the source being unassociated) yielded light-curves for 35 such unassociated sources.

We fit the light-curve of these sources with a simple model consisting of two parameters: the distance $d$ and the remaining PBH lifetime at initial detection, $\tau$. We utilize a simple parameterization derived from the approximate differential photon yield integrated over the remaining lifetime, here $\tau$, of an evaporating black hole. Calling $k_BT_\tau\approx 7.8\ {\rm TeV}\ (\tau/{\rm sec})^{-1/3}$ the temperature corresponding to the remaining lifetime $\tau$, Ref.~\cite{Bugaev:2007py, Petkov:2008rz} finds, on the basis of HERWIG-based Monte Carlo simulations of the photon yield, 
\begin{eqnarray}
    \frac{dN_\gamma}{dE_\gamma}&\approx& 9\times 10^{35}\left(\frac{1\ \rm GeV}{k_B T_\tau}\right)^{3/2}\left(\frac{\rm 1\ GeV}{E_\gamma}\right)^{3/2}\ {\rm GeV}^{-1}\quad (E_\gamma<k_BT_\tau)\\
    \frac{dN_\gamma}{dE_\gamma}&\approx& 9\times 10^{35} \left(\frac{\rm 1\ GeV}{E_\gamma}\right)^{3}\ {\rm GeV}^{-1}\quad (E_\gamma\ge k_BT_\tau).
\end{eqnarray}
Considering the integrated flux above 0.1 GeV, we find that the integrated flux from a source at a distance $d$ as a function of time from detection $t$ (and indicating with $\tau$ again the remaining PBH lifetime at initial detection) reads, approximately,
\begin{equation}\label{eq:fittingeq}
    F_\gamma(t,E_\gamma>0.1\ {\rm GeV})\simeq 2.7\times 10^{-8}\left(\frac{\rm pc}{d}\right)^2\left(\frac{\tau-t}{\rm sec}\right)^{-0.533}\ {\rm cm}^{-2}{\rm sec}^{-1}.
\end{equation}
Note that this is also consistent with Eq.~(\ref{eq:ngammadot}), when integrated over the same energy range. In what follows, we use Eq.~(\ref{eq:fittingeq}) to fit for the distance and age of long-term evaporating PBHs.

Since the source continues to emit radiation, we use the observation window $t_{\rm{obs}}$ as a lower-bound on $\tau>t_{\rm{obs}}$. This value is indicated in Fig.~\ref{fig:transients} as $\Delta MJD$, where MJD stands for ``MJD=Modified Julian Date = Julian Date - 2400000.5''.  We leave the remaining bounds for $d$ and $\tau$ open. We assume that  each error of each datapoint is Gaussian, and apply a non-linear least squares regression to each light-curve for the light-curve model of Eq.~(\ref{eq:fittingeq}). The fitting process leaves wide errors for the model, owing to the small dataset size. We   do not fit sources with a single datapoint, for which we independently calculate the standard error of the model  with the 50\% confidence intervals as shown. The fitted parameters for each source are given in Tab.~\ref{tab:longterm}, and the individual sources are shown in Fig.~\ref{fig:longsrc}. In the table, the columns correspond to the source name, right ascension (RA), declination (DEC), remaining time to complete evaporation $\tau$, with its upper and lower limits, fitted distance, with corresponding upper and lower limits, and the spectral index at 1 GeV, and its error. We can very clearly see the resulting effect of the errors when we plot, in Fig.~\ref{fig:longsrcfig} our results on the same plane as Fig.~\ref{fig:detsens}. 

We find that the sources, as expected, fall on the mass versus distance plane {\em below} the sensitivity predicted for the LAT, and they cluster into two groups: a first group, with masses around $10^{12}$ grams and distances of around a milli-parsec, which exhibit a relatively fast-increasing light-curve; and a second group, at masses between $10^{14}$ and $10^{15}$ grams and distances of $10^{-5}-10^{-4}$ pc, with a slowly increasing light-curve. The first set of sources are expected to have proper motion in the few degrees range, on average. We note that this is {\em not consistent} with the fact that the LAT Transient Catalog \cite{LATTransients} does not flag these sources as being associated with any proper motion. The second group of sources should have an even larger motion in the sky, several tens of degrees, again inconsistent with the LAT Transient Catalog. We also note that the sources in the first, low-mass group should be detectable with operating high-energy gamma-ray Cherenkov telescopes, but such detections are lacking. All sources, finally, should have been detectable by GBM and/or BATSE, but again they were not. We thus conclude that {\em circumstantial evidence is strongly against the association of these long-term transient sources with evaporating, nearby PBHs}.

An additional test, albeit inconclusive due to the current quality of data, of whether the sources under consideration can be associated with evaporating PBH is to inspect the gamma-ray spectrum. As a proxy thereof, we compare, in Fig.~\ref{fig:spectralindex}, the spectral index $\gamma$ at a gamma-ray energy of $E_0=1$ GeV, as defined in Eq.~(\ref{eq:spectralindex}) above, with the expected spectral index for evaporating PBHs shown in Fig.~\ref{fig:gammasp}, as a function of the PBH lifetime left at initial detection  $\tau$.
This quantity provides weak constraints due to the wide errors in the lifetime estimation. However, we note that the spectral indexes are generically {\em softer} than what expected for the light-mass cluster sources (i.e. $\gamma$ is larger than predicted), while for the higher-mass cluster (corresponding in Fig.~\ref{fig:spectralindex} to larger remaining lifetimes at detection, i.e. the left of the plot), the spectrum is inconsistent with the predicted exponential suppression, at least for the central values for $\tau$.


\begin{figure}
    \centering
    \includegraphics[width = \linewidth, trim={5cm, 7cm, 5cm, 7cm}, clip]{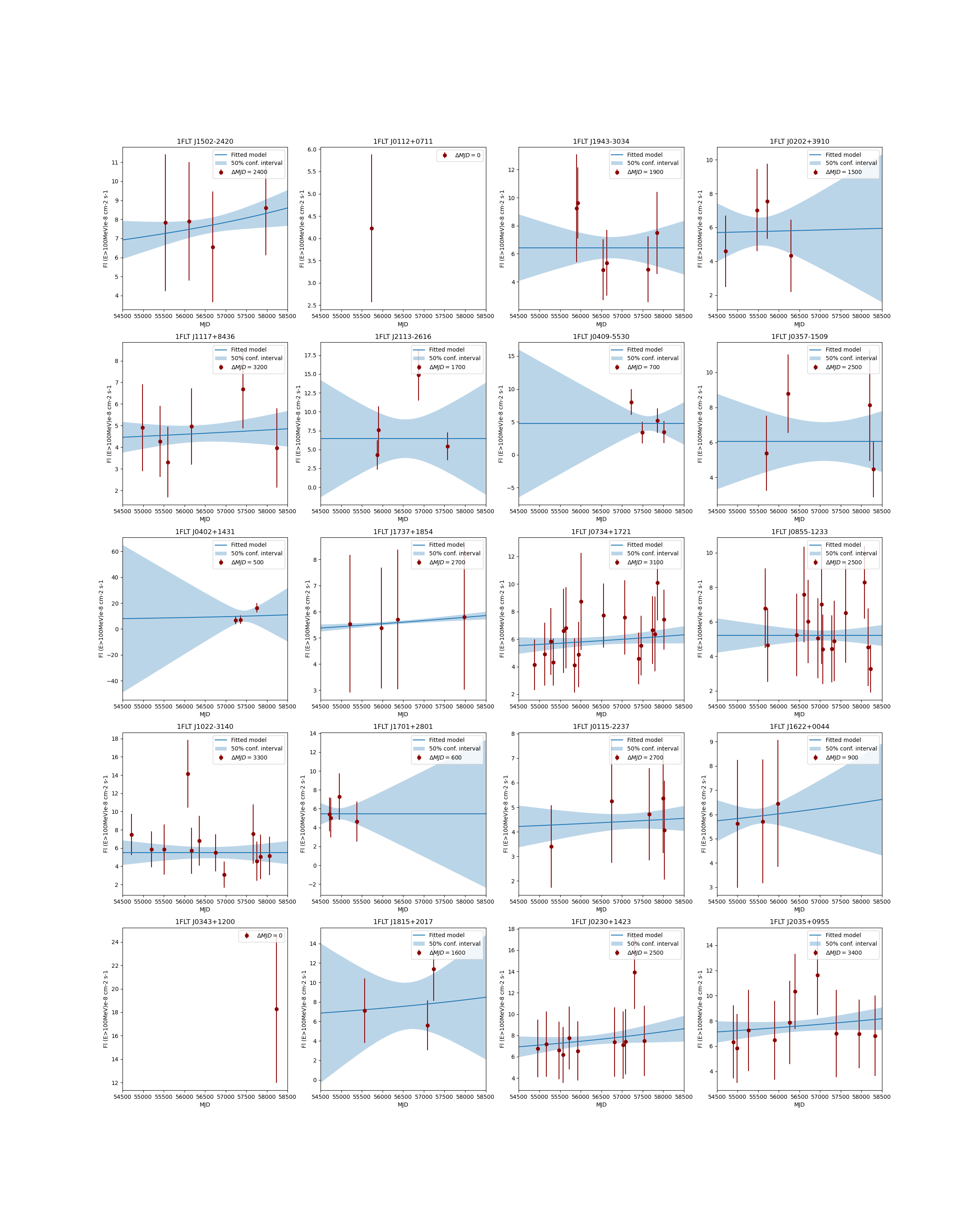}
\end{figure}

\begin{figure}
    \centering
    \includegraphics[width = \linewidth, trim={5cm, 4cm, 5cm, 7cm}, clip]{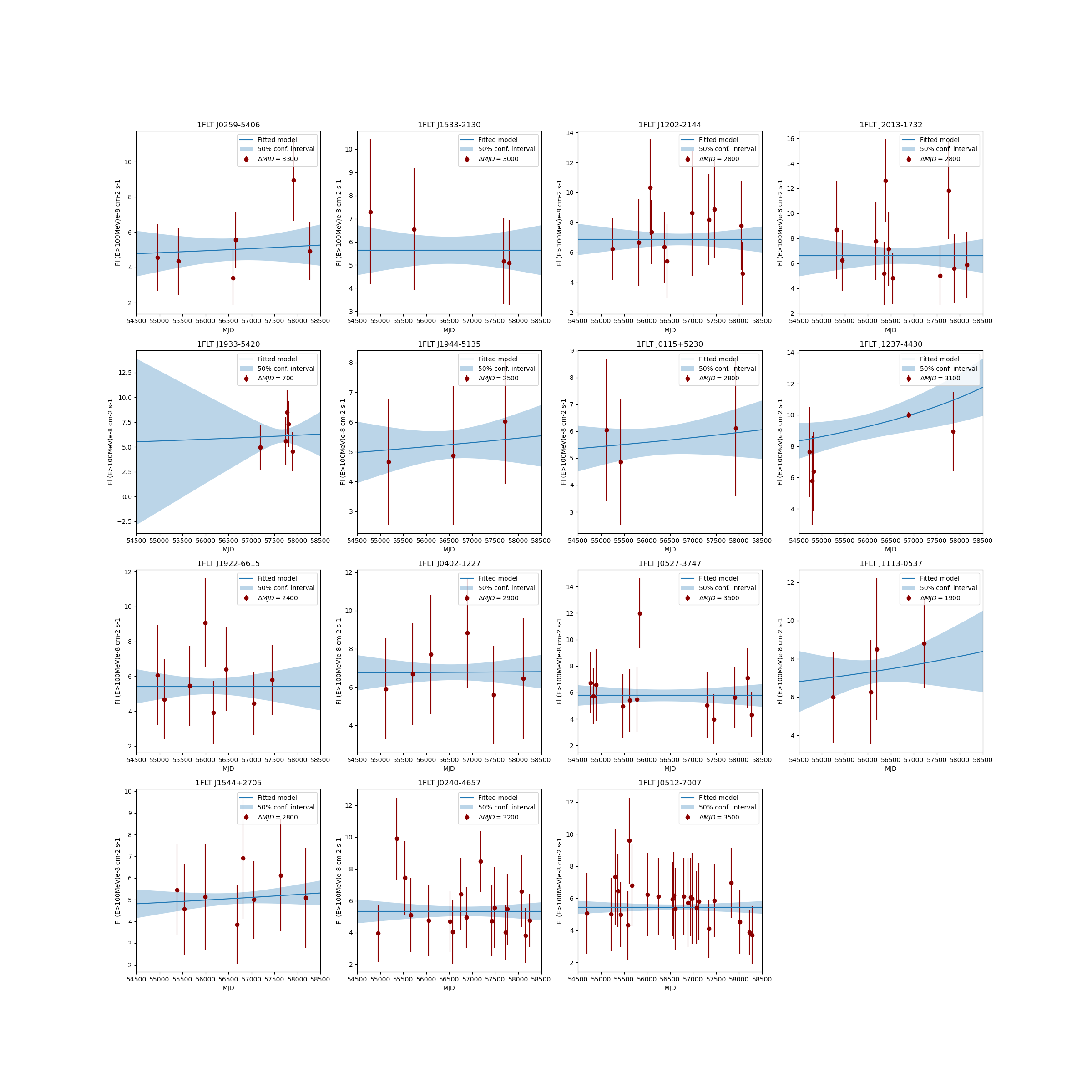}
    \caption{Parameterized power-law model (blue) with 50\% confidence interval (blue shaded) for unassociated gamma-ray transients detected by the \textit{Fermi}-LAT (red error-bars). Duration of the signal is shown for reference in Modified Julian Days, $T_{\rm{obs}} = \Delta MJD$.}
    \label{fig:longsrc}\label{fig:transients}
\end{figure}

\begin{figure}
\begin{minipage}{\linewidth}
    \centering
    \includegraphics[width=\linewidth]{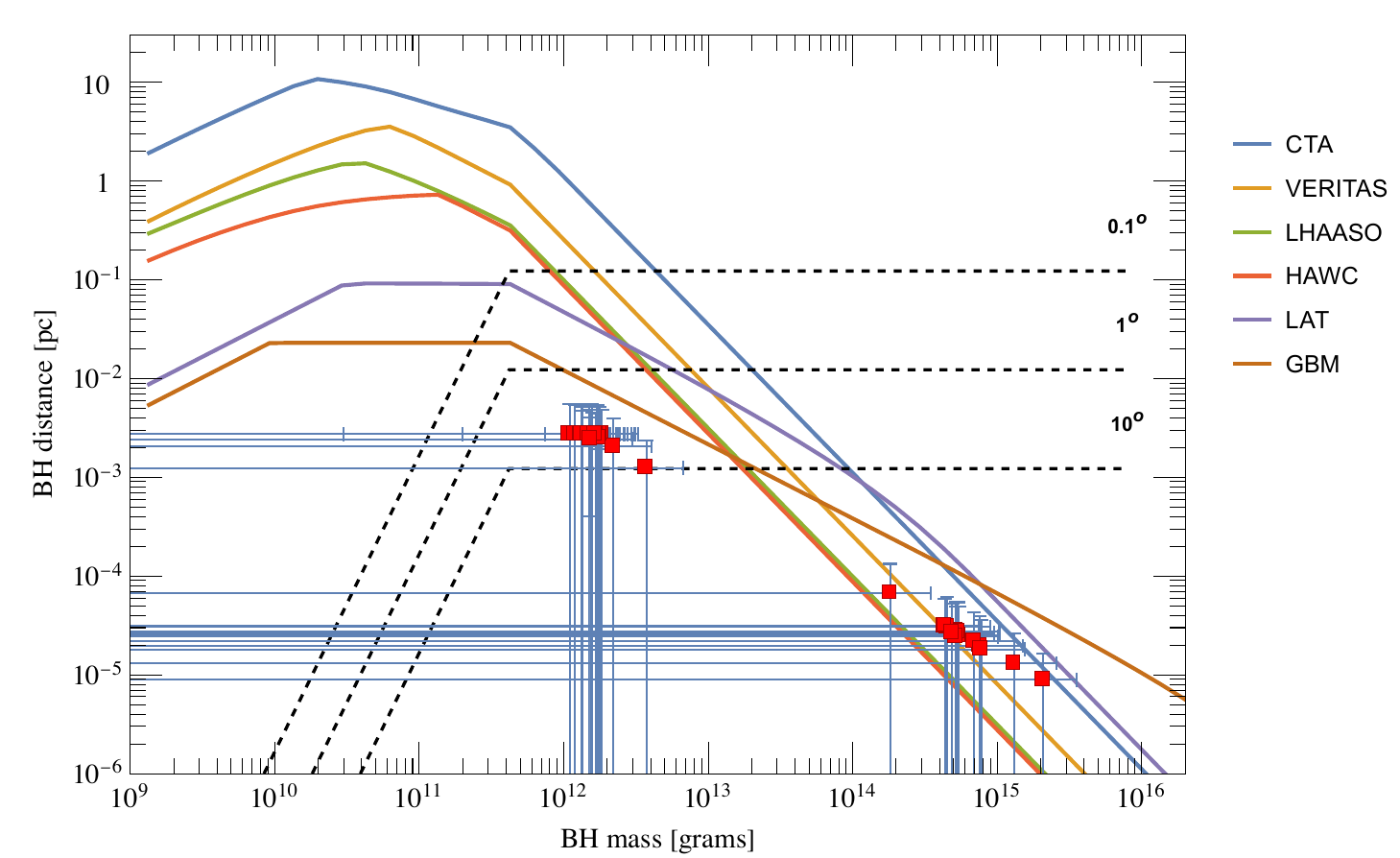}
    \caption{Transient fit results superimposed on the projected sensitivity limits described in previous sections.}\label{fig:longsrcfig}
\end{minipage}\\
\begin{minipage}{\linewidth}
    \centering
    \includegraphics[width=\linewidth]{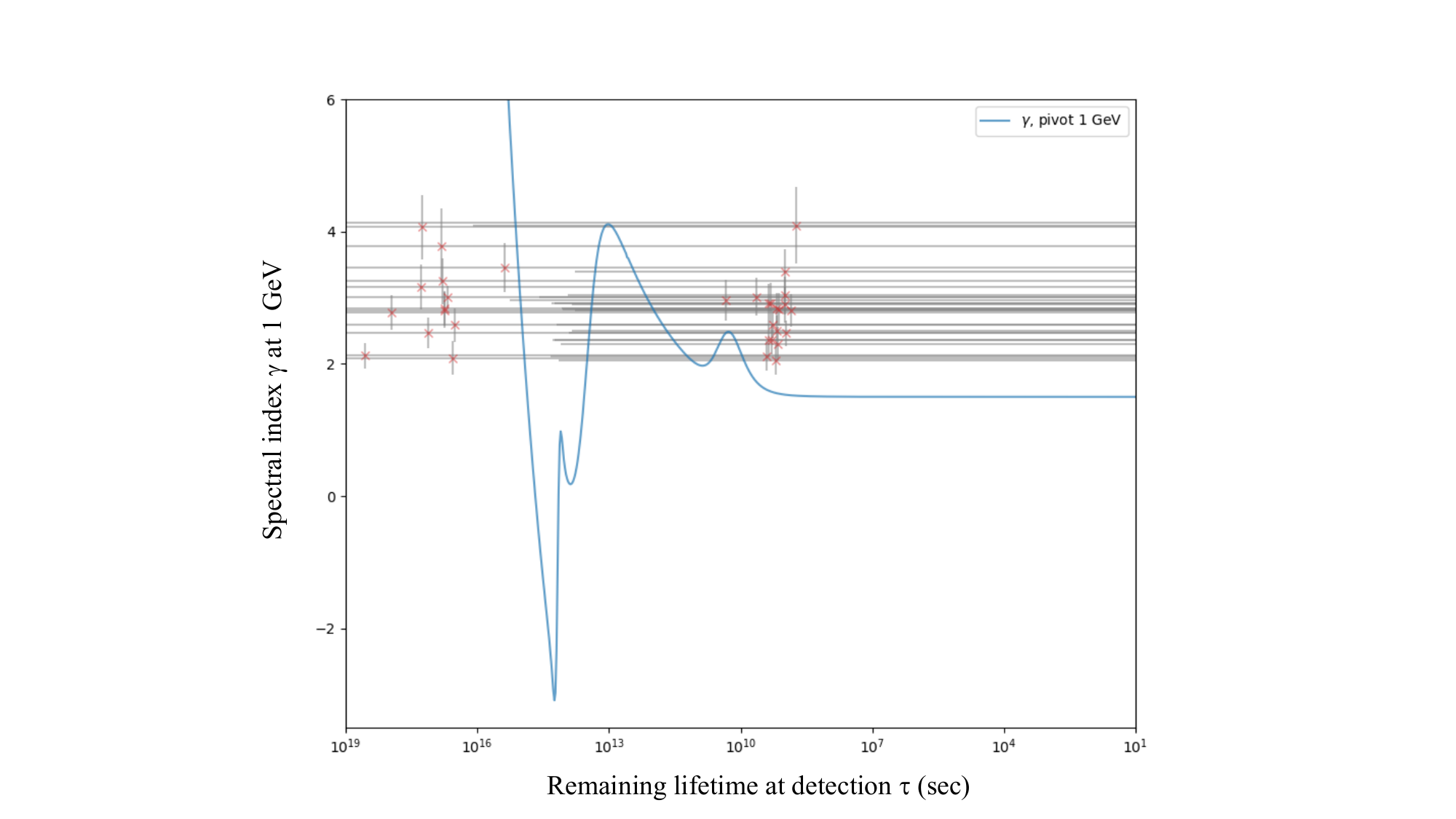}
    \caption{Lifetime and spectral index $\gamma$ as fitted for the transient sources compared to the theoretical expectations for the same.}\label{fig:spectralindex}
\end{minipage}
\end{figure}

\begin{landscape}
\begin{table}
{\small
\centering
\begin{tabular}{|l|l|l|l|l|l|l|l|l|c|c|}

\hline
Source name & RA (deg) & DEC (deg) & $\tau$ (s) & $\tau$ u.l. (s) & $\tau$ l.l. (s) & $d$ (pc) & $d$ u.l. (pc) & $d$ l.l. (pc) & $\gamma$ & $\gamma$ err.\\
\hline
1FLT J0115+5230	&	18.86	&	52.52	&	1.61E+09	&	1.25E+10	&	2.08E+08	&	2.74E-03	&	7.61E-03	&	9.83E-04	&	2.06	&	0.23	\\
1FLT J0115-2237	&	18.96	&	-22.63	&	2.55E+09	&	4.05E+10	&	1.60E+08	&	2.74E-03	&	1.09E-02	&	6.86E-04	&	2.12	&	0.22	\\
1FLT J0202+3910	&	30.60	&	39.17	&	4.41E+09	&	9.57E+11	&	2.03E+07	&	2.05E-03	&	3.02E-02	&	1.39E-04	&	3.01	&	0.29	\\
1FLT J0230+1423	&	37.64	&	14.39	&	9.84E+08	&	2.53E+13	&	3.83E+04	&	2.74E-03	&	4.39E-01	&	1.71E-05	&	3.39	&	0.35	\\
1FLT J0240-4657	&	40.16	&	-46.95	&	5.67E+16	&	3.22E+21	&	1.00E+12	&	2.77E-05	&	6.59E-03	&	1.16E-07	&	2.84	&	0.26	\\
1FLT J0259-5406	&	44.84	&	-54.10	&	2.03E+09	&	2.55E+12	&	1.61E+06	&	2.74E-03	&	9.71E-02	&	7.71E-05	&	2.92	&	0.30	\\
1FLT J0357-1509	&	59.40	&	-15.16	&	6.56E+16	&	8.76E+19	&	4.91E+13	&	2.50E-05	&	9.13E-04	&	6.84E-07	&	3.78	&	0.56	\\
1FLT J0402-1227	&	60.51	&	-12.46	&	2.13E+10	&	2.22E+12	&	2.05E+08	&	1.24E-03	&	1.26E-02	&	1.22E-04	&	2.97	&	0.31	\\
1FLT J0402+1431	&	60.66	&	14.52	&	5.39E+08	&	1.26E+16	&	2.30E+01	&	2.74E-03	&	1.32E+01	&	5.65E-07	&	4.10	&	0.59	\\
1FLT J0409-5530	&	62.29	&	-55.50	&	2.44E+15	&	3.06E+18	&	1.95E+12	&	6.74E-05	&	2.38E-03	&	1.90E-06	&	3.46	&	0.36	\\
1FLT J0512-7007	&	78.05	&	-70.13	&	9.21E+17	&	1.86E+22	&	4.55E+13	&	1.31E-05	&	1.86E-03	&	9.21E-08	&	2.77	&	0.27	\\
1FLT J0527-3747	&	81.80	&	-37.79	&	1.77E+17	&	1.35E+22	&	2.32E+12	&	1.97E-05	&	5.43E-03	&	7.13E-08	&	4.07	&	0.49	\\
1FLT J0734+1721	&	113.70	&	17.36	&	1.54E+09	&	6.17E+13	&	3.82E+04	&	2.74E-03	&	5.48E-01	&	1.36E-05	&	2.49	&	0.23	\\
1FLT J0855-1233	&	133.75	&	-12.56	&	6.23E+16	&	3.48E+20	&	1.12E+13	&	2.73E-05	&	2.04E-03	&	3.66E-07	&	3.25	&	0.35	\\
1FLT J1022-3140	&	155.74	&	-31.67	&	3.73E+16	&	1.93E+23	&	7.24E+09	&	3.05E-05	&	6.92E-02	&	1.34E-08	&	2.09	&	0.26	\\
1FLT J1113-0537	&	168.36	&	-5.63	&	9.96E+08	&	5.51E+10	&	1.80E+07	&	2.74E-03	&	2.03E-02	&	3.68E-04	&	3.04	&	0.28	\\
1FLT J1117+8436	&	169.28	&	84.62	&	2.31E+09	&	1.76E+11	&	3.02E+07	&	2.74E-03	&	2.39E-02	&	3.13E-04	&	2.92	&	0.29	\\
1FLT J1202-2144	&	180.73	&	-21.74	&	5.63E+16	&	4.02E+20	&	7.87E+12	&	2.44E-05	&	2.07E-03	&	2.89E-07	&	2.81	&	0.27	\\
1FLT J1237-4430	&	189.48	&	-44.51	&	7.03E+08	&	8.91E+11	&	5.55E+05	&	2.74E-03	&	9.74E-02	&	7.68E-05	&	2.81	&	0.25	\\
1FLT J1502-2420	&	225.73	&	-24.34	&	9.38E+08	&	1.26E+10	&	6.97E+07	&	2.74E-03	&	1.00E-02	&	7.46E-04	&	2.46	&	0.19	\\
1FLT J1533-2130	&	233.45	&	-21.50	&	3.67E+18	&	1.64E+20	&	8.21E+16	&	8.92E-06	&	5.96E-05	&	1.33E-06	&	2.12	&	0.19	\\
1FLT J1544+2705	&	236.04	&	27.10	&	1.96E+09	&	1.20E+11	&	3.21E+07	&	2.74E-03	&	2.14E-02	&	3.50E-04	&	2.37	&	0.22	\\
1FLT J1622+0044	&	245.71	&	0.74	&	1.42E+09	&	4.89E+09	&	4.13E+08	&	2.74E-03	&	5.07E-03	&	1.47E-03	&	2.30	&	0.25	\\
1FLT J1701+2801	&	255.42	&	28.02	&	3.36E+16	&	1.54E+18	&	7.30E+14	&	3.14E-05	&	2.13E-04	&	4.63E-06	&	2.59	&	0.25	\\
1FLT J1737+1854	&	264.27	&	18.90	&	2.27E+09	&	3.43E+09	&	1.50E+09	&	2.50E-03	&	3.07E-03	&	2.03E-03	&	2.36	&	0.27	\\
1FLT J1815+2017	&	273.95	&	20.29	&	9.53E+08	&	1.34E+13	&	6.79E+04	&	2.74E-03	&	3.24E-01	&	2.31E-05	&	2.90	&	0.27	\\
1FLT J1922-6615	&	290.61	&	-66.25	&	1.34E+17	&	1.66E+20	&	1.09E+14	&	2.18E-05	&	7.67E-04	&	6.22E-07	&	2.47	&	0.24	\\
1FLT J1933-5420	&	293.42	&	-54.34	&	1.34E+09	&	4.59E+11	&	3.91E+06	&	2.74E-03	&	5.06E-02	&	1.48E-04	&	2.83	&	0.25	\\
1FLT J1943-3034	&	295.91	&	-30.57	&	4.74E+16	&	2.47E+20	&	9.09E+12	&	2.64E-05	&	1.91E-03	&	3.66E-07	&	3.01	&	0.17	\\
1FLT J1944-5135	&	296.12	&	-51.60	&	1.85E+09	&	1.23E+10	&	2.79E+08	&	2.74E-03	&	7.04E-03	&	1.06E-03	&	2.59	&	0.20	\\
1FLT J2013-1732	&	303.50	&	-17.55	&	1.91E+17	&	3.45E+23	&	1.05E+11	&	1.81E-05	&	2.43E-02	&	1.34E-08	&	3.16	&	0.34	\\
1FLT J2035+0955	&	308.90	&	9.92	&	1.48E+09	&	1.16E+13	&	1.89E+05	&	2.43E-03	&	2.15E-01	&	2.75E-05	&	2.84	&	0.24	\\
1FLT J2113-2616	&	318.37	&	-26.27	&	4.58E+27	&	7.67E+34	&	2.73E+20	&	3.24E-08	&	1.33E-04	&	7.93E-12	&	4.14	&	0.58	\\
\hline
\end{tabular}
}
\caption{Gamma-ray LAT transients, also shown in Fig.~\ref{fig:longsrc}, with relevant statistics and fitted parameters. The columns correspond to the source name, right ascension (RA), declination (DEC), remaining time to complete evaporation $\tau$, with its upper and lower limits, fitted distance, with corresponding upper and lower limits, and the spectral index at 1 GeV, and its error.}\label{tab:longterm}
\end{table}
\end{landscape}

\subsection{Short-duration GRB-like sources}
\begin{figure}
\begin{minipage}{.4\linewidth}
    \centering
    \hspace*{-0.5cm}\includegraphics[width=1.1\linewidth]{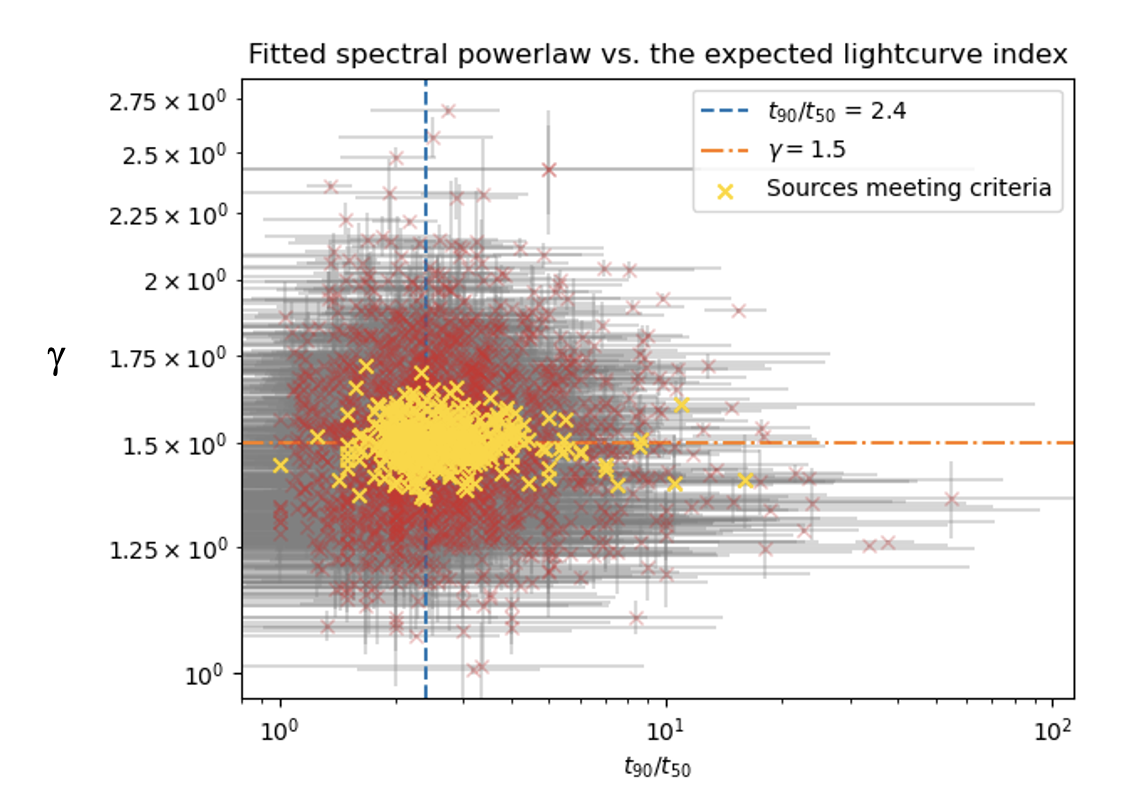}
    \caption{Candidate GRB events possibly compatible with PBH evaporation, on the $(t_{90}/t_{50},\gamma)$ plane.}
    \label{fig:GRB}
\end{minipage}\hfill
\begin{minipage}{.55\linewidth}
    \centering
    \includegraphics[width = \linewidth]{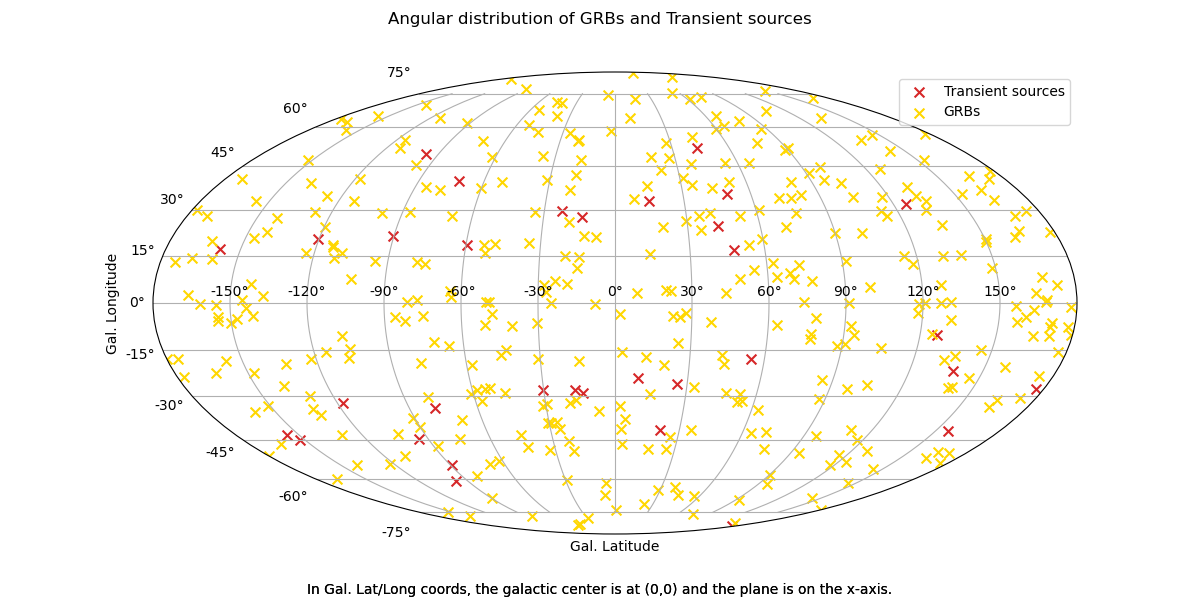}
    \caption{Angular distribution of PBH-candidate GRBs and Transient gamma-ray sources as catalogued in the GBM and Transient source catalogs.}
    \label{fig:my_label}
\end{minipage}
\end{figure}

In searching for events potentially associated with the last phase of PBH evaporation, we use the Fermi GBM Burst Catalog, which lists high-energy sources detected in 14 years of operation \cite{GBMCat1, GBMCat2, GBMCat3, GBMCat4}. We impose the following criteria on the light-curve evolution and on the spectrum:
\begin{enumerate}
    \item In the GBM energy range, the expectation for the ratio $t_{90}/t_{50} \simeq 2.4$, which is a proxy for the light-curve evolution with time for exploding PBHs (here $t_{50}$ and $t_{90}$ indicate the time taken to accumulate 50\% (90\%) of the burst fluence starting at the 25\% (respectively, 5\%) fluence level)
    \item For a pivot energy below the GeV, the spectral index is expected to follow $d\phi_\gamma/dE\sim E^{-\gamma_{\rm PBH}}$ with $\gamma_{\rm PBH}\simeq1.5$, the late-time asymptote shown in Fig.~\ref{fig:gammasp}.
\end{enumerate}
For both quantities, we utilize the values of $t_{90}$, $t_{50}$, and $\gamma$ listed in the GBM event catalogue. We include all sources that are 2$\sigma$ within the predicted values of $t_{90}/t_{50}$ and $\gamma_{\rm PBH}$. The resulting list of candidate sources amounts to the 388 shown in Fig.~\ref{fig:GRB}, left, with yellow x's, on the plane defined by the spectral index versus the $t_{90}/t_{50}$ ratio (the vertical and horizontal lines indicate the PBH expected values).

Fig.~\ref{fig:GRB}, right, shows the sky distribution of both the GRB events (again indicated as yellow x's) and the unassociated transient sources discussed above. We note a fairly homogeneous sky distribution for both classes of events. In particular, we find that the ratio of sources in the northern to southern hemisphere is 1.04, whereas the ratio of sources in the western to eastern hemispheres is 0.879 -- indicating a slightly larger abundance in the eastern hemisphere of the sky. The theoretical expectation for PBH events, stemming from the relative position of the Earth within the Milky Way dark matter halo in Galactic coordinates, and assuming no PBH clustering \cite{Gorton:2022fyb} is of order 0.02\% for a sphere of radius 1 pc and 0.2\% for a radius of 10 pc, assuming a standard Navarro-Frenk-White \cite{Navarro:1995iw} dark matter density profile, i.e., indicating with $N_{\rm GC}(R)$ the expected number of events from an hemisphere of radius $R$ centered on the Earth at pointed towards the Galactic center  
and with $N_{\rm AGC}(R)=N_{\rm tot}(R)-N_{\rm GC}(R)$ those expected from the opposite hemisphere in the direction of the anti-Galactic center, $(N_{\rm GC}-N_{\rm AGC})/N_{\rm tot}(1\ {\rm pc})\simeq 2\times 10^{-4}$ and $(N_{\rm GC}-N_{\rm AGC})/N_{\rm tot}(10\ {\rm pc})\simeq 2\times 10^{-3}$;
Note that because of the spherical symmetry of the Navarro-Frenk-White profile, no asymmetry is expected in the east vs west hemispheres. We note that previous studies of very-short GRBs using the BATSE monitor discovered a similar overdensity in the eastern hemisphere, caused possibly by halo clumping \cite{Cline_2003}. Note also that isotropy is an indication of extragalactic gamma-ray sources, disfavoring, but not ruling out, a Galactic origin such as PBH explosions. Possible future directions in this area include pursuing solar neighborhood-localized bursts utilizing a network of GRB detectors, as suggested in \cite{Ukwatta:2015mfb}.

\subsection{Possible association of long-term transient and GRB sources with evaporating PBHs}

\begin{figure}
    \centering
    \includegraphics[width = \linewidth]{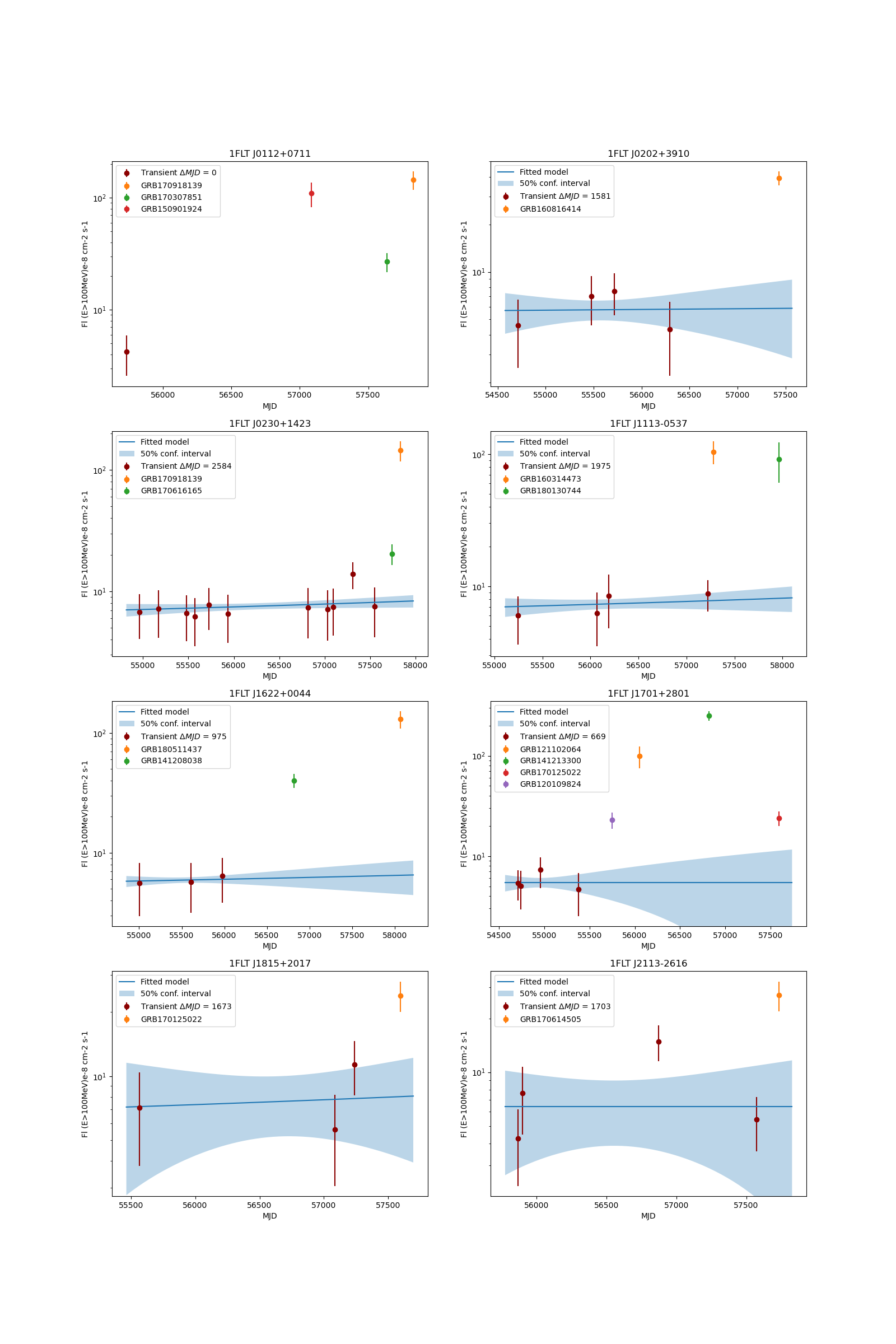}
    \vspace*{-2.5cm}
    \caption{For each LAT transient source, we plot the flux of tentatively associated GRBs. 
    }
    \label{fig:association}
\end{figure}

Here, we search for events where a long-term, unassociated, transient source from the LAT catalogue discussed in sec.~\ref{sec:longsrc} could be associated with a gamma-ray burst, accounting for proper motion. 
We consider any event where the difference between the LAT and GRB localization for right ascension (RA) and declination (DEC), accounting for uncertainties in the angular determination $\Delta \theta$, and allowing for 1 degree of proper motion per year (corresponding, approximately, to sources within 0.01 pc, the typical maximal distance for detection by the LAT) scaled by time (the initial detection time for the transient source and $t_{0}^{\rm{trans.}}$ and the GRB burst time $t^{\rm{grb}}$).
\begin{equation}
\begin{gathered}
    \sqrt{\left( \rm{RA}^{\rm{grb}} - \rm{RA}^{\rm{trans.}} \right)^2 + \left( \rm{DEC}^{\rm{grb}} - \rm{DEC}^{\rm{trans.}} \right)^2} - \left( \Delta \theta^{\rm{grb}} + \Delta \theta^{\rm{trans.}} \right) \\
    \leq \frac{1 \rm{deg}}{\rm{yr}} \times \left( t^{\rm{grb}} - t_{0}^{\rm{trans.}} \right)
\end{gathered}
\end{equation}
Of the 35 transient sources, we find 14 which have at least one GRB detected within the above limit. 
In an ideal case, the GRB follows the last (most recent) observation of a transient source and appears brighter than its transient counterpart as it corresponds with the final, runaway PBH explosion. Fig.~\ref{fig:association} shows the light-curves of the possible associations, in conjunction with the tentatively associated GRB events. The relevant catalog information and data are listed in Table \ref{tab:associations}, where we indicate, for each source, the source identification name, RA, DEC, radial extent error, flux and its error, ratio of $t_{90}/t_{50}$ and its error, MJD, $t_{90}$ and corresponding error, spectral index and its upper and lower limit. 
%
%
For all the fitted transient sources, the corresponding GRBs lie well outside the confidence interval to be  associated with the standard PBH light-curve, albeit in some cases the association is not obviously ruled out with large statistical significance given the error band of the long-term transient light-curve (for instance in the case of 1FLT J1701+2801 and GRB170125022).

\begin{figure}[!t]
    \centering
    \includegraphics[width=.7\linewidth]{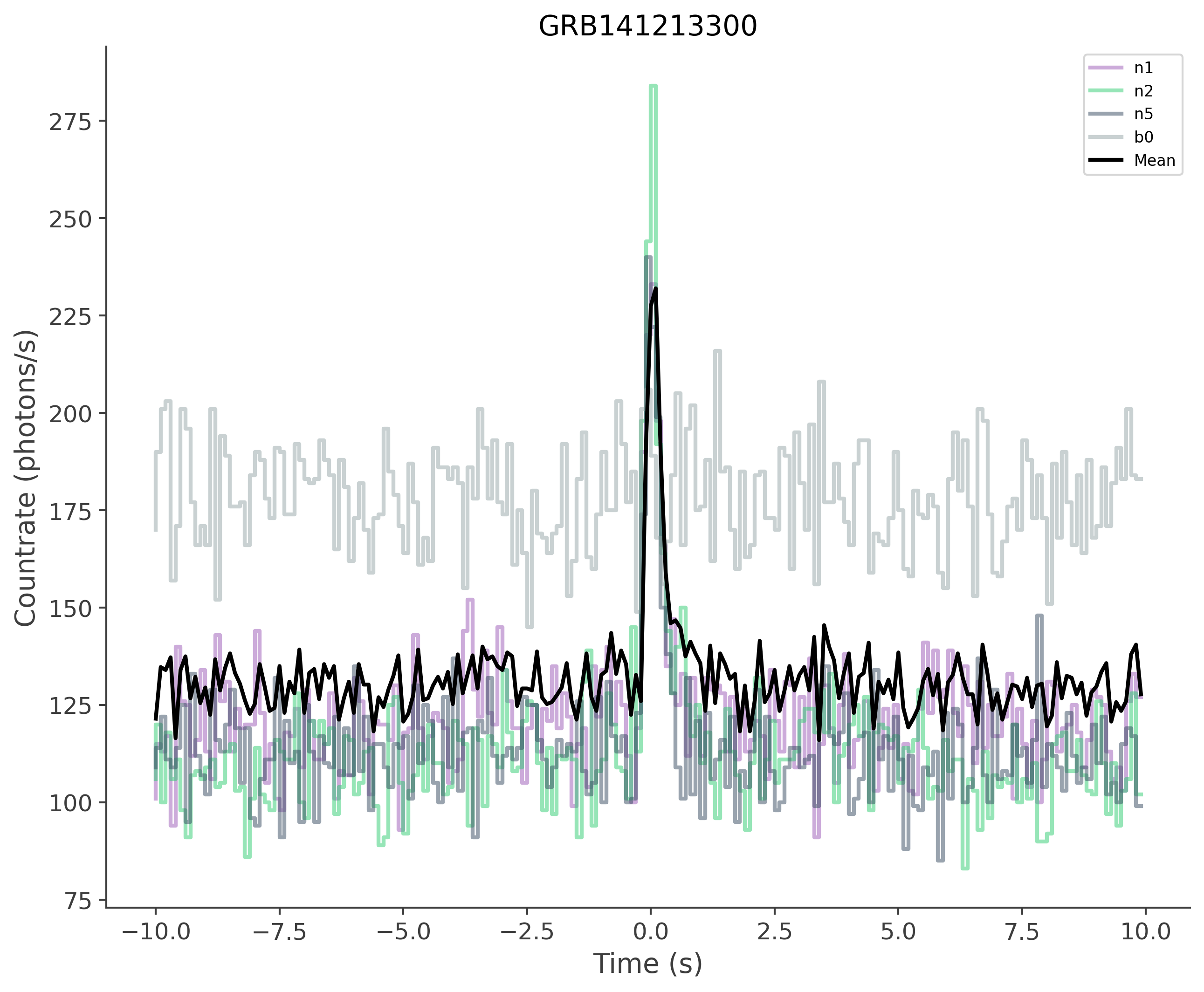}
    \caption{The light-curve of GRB141213300, for the four GBM detectors, and the corresponding mean (black); the event  lies within the scaled ROI of 1FLT J1701+2801. }
    \label{fig:GRB141213300}
\end{figure}

We additionally inspected the light-curves of all {\em individual} GRBs listed in Tab.~\ref{tab:associations}. In most cases, the presence of an ``afterglow'' strongly rules out the association of the event with the last stage of an evaporating PBH: as shown conclusively in Ref.~\cite{MacGibbon:2007yq}, no ``photosphere'', and therefore no afterglow can form around an evaporating PBH. Of all sources, the one GRB that, statistically, could be perhaps marginally compatible with the light-curve of an evaporating PBH is GRB141213300, shown in Fig.~\ref{fig:GRB141213300}. We show there the signal from the four GBM detectors, the NaI (colored n1, n2, n5) and BGO (grey b0) detectors, binned on 100ms and overlain with their average (black). This source typifies many of the ideal characteristics in a candidate PBH GRB, including sharp rising and falling edges. Additionally, the higher-energy BGO detector has a slightly higher count-rate and leads the NaI detectors by $\sim100$ms.  Intriguingly, GRB141213300 lies within the scaled ROI of 1FLT J1701+2801 (see Fig.~\ref{fig:association}), even though the measured flux of GRB141213300 lies well outside the possible extrapolated behavior of the light-curve of 1FLT J1701+2801, making the identification implausible at best. 


\begin{table}\label{tab:associations}
{\footnotesize
\centering
\begin{tabular}{|c|c|c|c|c|c|c|c|}
\hline
Source name & RA (deg) & DEC (deg) & Rad. Err. (deg) & Flux (erg/cm$^2$/s) & Flux err. & $t_{90}/t_{50}$ & $t_{90}/t_{50}$ err.\\
\hline
GRB170614505 & 310.99 & -37.91 & 15.16 & 2.71e-07 & 5.19e-08 & 2.10 & 1.62 \\ GRB160314473 & 161.99 & 2.83 & 19.99 & 1.04e-06 & 2.04e-07 & 8.67 & 15.87 \\ GRB170918139 & 36.56 & 3.52 & 17.81 & 1.45e-06 & 2.67e-07 & 4.00 & 6.96 \\ GRB180511437 & 257.78 & 9.07 & 10.16 & 1.30e-06 & 2.09e-07 & 2.58 & 1.78 \\ GRB170307851 & 13.54 & 9.54 & 0.05 & 2.69e-07 & 5.10e-08 & 2.06 & 0.79 \\ GRB141208038 & 239.16 & 10.97 & 9.49 & 4.01e-07 & 5.41e-08 & 1.93 & 0.57 \\ GRB150901924 & 16.34 & 13.52 & 17.12 & 1.09e-06 & 2.73e-07 & 2.00 & 9.29 \\ GRB121102064 & 258.47 & 14.09 & 12.15 & 9.90e-07 & 2.45e-07 & 1.60 & 1.52 \\ GRB141213300 & 248.19 & 18.06 & 8.72 & 2.49e-06 & 2.72e-07 & 4.00 & 3.62 \\ GRB170616165 & 49.51 & 19.67 & 12.12 & 2.04e-07 & 3.93e-08 & 2.27 & 0.74 \\ GRB170125022 & 264.14 & 28.58 & 12.65 & 2.40e-07 & 3.93e-08 & 2.44 & 1.49 \\ GRB120109824 & 251.33 & 30.80 & 11.33 & 2.29e-07 & 4.15e-08 & 2.25 & 0.72 \\ GRB160816414 & 25.32 & 43.70 & 19.13 & 3.93e-07 & 3.94e-08 & 2.19 & 1.53 \\ GRB180130744 & 136.83 & 52.69 & 68.08 & 9.20e-07 & 3.14e-07 & 2.00 & 7.58 \\
\hline
\end{tabular}
\vspace{3em}

\begin{tabular}{|c|c|c|c|c|c|c|}
\hline
Source name & MJD (days)  & $t_{90}$ (s) & $t_{90}$ err. & $\gamma$ & $\gamma$ pos. err. & $\gamma$ neg. err.\\
\hline
GRB170614505 & 5.77e+04 & 5.38 & 1.64 & -1.45 & 0.12 & 0.12 \\ GRB160314473 & 5.73e+04 & 1.66 & 0.73 & -1.51 & 0.13 & 0.13 \\ GRB170918139 & 5.78e+04 & 0.13 & 0.16 & -1.45 & 0.08 & 0.08 \\ GRB180511437 & 5.81e+04 & 1.98 & 0.97 & -1.52 & 0.10 & 0.10 \\ GRB170307851 & 5.76e+04 & 28.42 & 1.72 & -1.57 & 0.10 & 0.10 \\ GRB141208038 & 5.68e+04 & 14.34 & 1.45 & -1.49 & 0.07 & 0.07 \\ GRB150901924 & 5.71e+04 & 0.26 & 1.15 & -1.40 & 0.17 & 0.17 \\ GRB121102064 & 5.60e+04 & 2.05 & 1.38 & -1.52 & 0.16 & 0.16 \\ GRB141213300 & 5.68e+04 & 0.77 & 0.51 & -1.54 & 0.05 & 0.05 \\ GRB170616165 & 5.77e+04 & 56.32 & 6.08 & -1.39 & 0.12 & 0.12 \\ GRB170125022 & 5.76e+04 & 3.90 & 1.12 & -1.45 & 0.10 & 0.10 \\ GRB120109824 & 5.58e+04 & 38.66 & 3.11 & -1.50 & 0.11 & 0.11 \\ GRB160816414 & 5.74e+04 & 11.78 & 3.81 & -1.54 & 0.06 & 0.06 \\ GRB180130744 & 5.80e+04 & 0.26 & 0.92 & -1.53 & 0.23 & 0.23 \\ 
\hline
\end{tabular}

}
\caption{Data and fitting parameters for the GRB sources  in Fig.~\ref{fig:association}, with relevant statistics. Note that MJD is Modified Julian Days. The various columns correspond to the source identification name, RA, DEC, radial extent error, flux and its error, ratio of $t_{90}/t_{50}$ and its error, MJD, $t_{90}$ and corresponding error, spectral index and its upper and lower limit.}
\end{table}

\section{Discussion, conclusions, and further directions}\label{sec:discussion}
The potential of directly observing a PBH explosion carries far-reaching implications for our understanding of the universe, from cosmology, to quantum gravity, to particle physics. While the existence of PBHs as a candidate for dark matter has been theorized for decades, direct detection of an evaporating PBH would provide invaluable insights into yet-undiscovered high-energy particles and dark radiation.

In this study, we have reviewed photon emission from the late-stage evaporation of PBHs. We produced a compact expression for the rate of evaporation of PBH as a function of their time-evolved mass function, and obtained upper limits on the local rate of PBH evaporation for several different mass functions. We showed that it is possible, albeit not expected, to have evaporation rates as large as the current observational limits from the HAWC telescope; more plausible rates hover around a few per tens of parsec cubed per year.

We then built an updated estimate of detector sensitivities across detectors as small as the Gamma-ray burst monitor onboard the Fermi satellite, and as large as LHAASO. We  computed the maximal distance where an exploding PBH with a given lifetime to expiration could be detected by a given telescope. We showed that the peak sensitivity is reached for lifetimes on the order of 100 to 10$^7$ sec, corresponding to masses in the ton to several hundred tons. We showed how larger detectors such as VERITAS and HAWC, and future detectors such as CTA and LHAASO are best suited to detect lighter PBHs (i.e. PBHs at later evaporation stages), while detectors such as LAT and GBM are best suited to detect heavier, and closer-by PBHs. We also calculated projected constraints on the evaporation rate as a function of the BH mass at the beginning of observations. We also studied the features of gamma-ray signals from PBH evaporation such as the spectral power-law index, that could help disentangle astrophysical, standard sources from PBHs and we evaluated the importance of source proper motion.

Sec.~\ref{sec:grb} was devoted to actual searches for exploding PBH in gamma-ray data from the Fermi satellite. We first focused on long-term variable, unassociated, ``transient'' sources from the LAT Transient Catalog. We individually fit the light-curves of the transient sources, and computed best-fit values for both the distance and the remaining PBH lifetime at initial detection; the sources cluster at extrapolated distances of (i) a few milli-parsec and lifetimes of 10$^8$-$10^9$ sec, corresponding to masses around $10^{12}$ grams, or at (ii) distances around a few tens of micro-parsecs, and lifetimes on the order of 10$^{16}$-$10^{19}$ sec, corresponding to masses in the $10^{14}$ to $10^{15}$ grams range; the two clusters correspond to ``fast rising'' vs slow rising'' transients. We noticed how our fits, however, have very large uncertainty in both distance and time to evaporation/mass. Additionally, the sources, if indeed associated with local evaporating PBHs should generally have significant proper motion, which they don't, and should have been observed by other observatories besides the LAT, which they haven't. We also studied the spectral properties of such long-term transients, and found that, even though, once again, the uncertainties are very significant, the spectra are not {\em per se} incompatible with what expected from PBH evaporation, especially for the fast rising transient sources.

We then explored short-duration GRB-like sources, and selected, amongst the Fermi GBM Burst Catalog, sources with the correct light-curve proxy, $t_{90}/t_{50}$, and spectral power-law index. We singled out close to 400 candidate sources, distributed roughly homogeneously in the sky. Further association of such searches with PBH evaporation events requires closer spectral inspection, and the possible association with long-term transient. Taking into account proper motion, we extracted a set of 8 possible associated long- and short-term sources; of these, one appears to have a light-curve roughly compatible with the absence of an afterglow (within statistical uncertainties). However, while our analysis did not produce any conclusive evidence of such association, it offers a framework for future studies and observations.

Overall, the potential for direct detection of a PBH explosion holds immense promise for advancing our understanding of the universe. As observational and theoretical analysis capabilities continue to improve, further investigation and refinements of the constraints presented in this study will undoubtedly lead to new insights in the physics of evaporating black holes.

\section*{Acknowledgments}

This work was supported in part by the U.S. Department of Energy grant number de-sc0010107 (SP).

\bibliographystyle{JHEP}
\bibliography{main}

\end{document}